\documentclass{article}

\def\be{\begin{equation}}
\def\ee{\end{equation}}
\def\bea{\begin{eqnarray}}
\def\eea{\end{eqnarray}}
\def\ben{\begin{eqnarray*}}
\def\een{\end{eqnarray*}}

\def\p{{\not\! p}}
\def\k{{\not\! k}}
\def\d{\partial}
\def\e{\varepsilon}
\def\<{\langle}
\def\>{\rangle}

\def\inp#1#2{\vec #1\! \cdot \! \vec #2}
\def\dpi{\vec\nabla\!\cdot\vec\pi}
\def\Dpi{\inp D \pi}		
\def\invHE{{1\over H_0-E - i\epsilon}}			

\def\delete#1{}

\begin{document}

\title{
	Non-fixing Gauge Field Quantization}

\author{Tsuguo MOGAMI
	\thanks{e-mail: mogami@brain.riken.go.jp}\\
	RIKEN, Hirosawa 2-1, Wako-shi,
	Saitama, 351-0198 Japan \\
}

\date{September 23, 2009}

\maketitle

\begin{abstract}
The gauge field theories are usually quantized by fixing gauge.
In this paper, we propose a new formalism that quantizes gauge fields  without gauge fixing but naturally follows canonical formalism.
New physical implications will follow.
\end{abstract}

Gauge symmetry is often indispensable because relativistic system usually need gauge symmetry.  If it were not for gauge symmetry for a vector or tensor field, we would have negative norm states when quantized.  Therefore gauge symmetry is requisite.

The problem is that a system with gauge symmetries is a ``constrained system", in which ordinary canonical formalism does not work.
Usually, it is considered necessary to eliminate freedom of gauge by hand since it is the only known way.
This is called ``gauge fixing".

In this paper, we propose a new formalism that quantizes gauge fields without gauge fixing but naturally follows canonical formalism.
Its main consequence is that this theory has quantized physical degree of freedom that corresponds to electrostatic field.
It will give the same result for scattering as the conventional theories, which is a desirable property.

In this paper we use the word ``electrostatic" in a little bit inexact sense.
It rather means longitudinal part of electric field $\vec E$.
It is not completely static as it changes as the charges move, but it is not dynamic like the transverse parts, which can propagate the space owing to the kinetic term in the Lagrangian.

The belief of transversality is that only transverse polarizations are physical degrees of freedom and taking transverse polarizations and quantizing it gives us the correct quantized field theory.
This belief is so ubiquitous that sometimes it is left unmentioned or even unnoticed.
\delete{FP determinant,} Feynman's ghost, BRST formalism, elimination of surplus degree of freedom follows and are equivalent to this.

The discourse in this paper will urge reconsideration to the belief of transversality, then all the works that rely on it will need rethink.
Further, after reading this paper, you will forget that you never knew about the static electric field in quantum field theories.

In section 1, QED will be quantized and we go over the canonical theory of constrained systems.
Then we can explicitly write down states of electrostatic field and calculate electrostatic energy.
In section 2, we apply our formalism to a non-Abelian gauge theory.
In these sections, remarks about the conventional theories were put into the footnotes because the author believes that these are not important for the future readers.
In section 3, we will review the conventional theories of quantizing gauge theories, and show that these all lack electrostatic field and equivalent to each other.

Conventions:			\break
In this paper, the vibration mode or polarization $A^\mu \propto (0, \vec k)$ is called ``L mode" or ``L polarization", where ``L" is for longitudinal.
The reason why we don't call it ``longitudinal polarization" is we want to avid confusion since sometimes people call $A^\mu \propto (k_0, \vec k)$ or $A^\mu \propto (|\vec k|, \vec k)$ as ``longitudinal polarization".

The sign of metric  $g_{\mu\nu}$ is \[
	(+,-,-,-).
\]
Three vectors are 3d spatial component of contravariant vectors, like
\[
	k^\mu = (k^0, \vec k).
\]
Exception is that $\d ^\mu = (\d^0, -\vec \nabla )$, because $\vec \nabla $ is defined as $\d /\d \vec x$.
We denote 4-vectors by plain letters like $x, p$ and their inner product without centered dot like $ e^{-i px}$.  A 3-vector will be signified by an arrow on it. 
Greek indices of the 4-vectors, e.g.\ $k_\mu, k^\mu$, run from 0 to 3.
Roman indices run over spatial component, i.e.\ 1,2,3.
Therefore $k_\mu x^\mu = k_0 x_0 - k_i x_i = k^0 x^0 - k^i x^i$.

$\omega$ denotes a infinitesimal fictitious mass of the gauge boson.
$ H$ denotes Hamiltonian, whose content differs from a section to another, and $H_0$ is its free part and $V$ is its interaction part.
Charge density is denoted by $\rho = e j_0$.

\section{QED}
The Lagrangian for quantum electrodynamics (QED) is \be
		L = \int d^3x \Big\{ - {1\over 4} (F_{\mu\nu})^2 + e j^\mu A_\mu \Big\},
	\label{LQED}
\ee
where $F_{\mu\nu}= \d_\mu A_\nu - \d_\nu A_\mu$.  This Lagrangian has gauge invariance.

Canonical quantization is converting classical Lagrangian formalism into classical canonical (Hamiltonian) formalism and then replacing its Poisson brackets by commutation relations.
However, QED cannot be converted into canonical formalism, because it is a so-called ``constrained system", which is inevitable for a system with any gauge invariance.

\subsection{Hamilton formalism of classical constrained systems}

Here we introduce canonical formalism for QED, which is a constrained system.  We first define conjugate momentum for the field $A_\mu$ as \be
	\pi^\mu = \frac{\d L}{\d\dot A_\mu},			\label{3}
\ee
and then introduce the Hamiltonian \be
	H(A, \pi) = \int d^3x\; \pi^\mu(x) \dot A_\mu(x) - L ,			\label{2}
\ee
and solve eq.(\ref{3}) to write $\dot A_\mu$ as functions of $\pi^\mu$ and $A_\mu$, and then we can write $H$ as a function of $A_\mu$ and $\pi^\mu$ only.

In the case of the gauge theory, we cannot express $\dot A_0$ in terms of $\pi^\mu$ and $A_\mu$, since $\pi_0 = 0$.  When we can't move some of the canonical variables independently, such limitations are called ``constraints" and such a system is called ``a constrained system".
We will write $\pi_0 = 0$ in a generalized form $ \phi(\pi_\mu, A_\mu) = 0$, which shows us the generalized canonical formalism of constrained systems.

By using eq.(\ref{3}), variation of $H$ is \ben
	\delta H &=& (\delta\pi^\mu) \dot A_\mu +\pi^\mu \delta\dot A_\mu
		- {\d L\over \d\dot A_\mu} \delta\dot A_\mu - {\d L\over \d A_\mu} \delta A_\mu			\\
		&=& \dot A_\mu \, \delta\pi^\mu  - {\d L\over \d A_\mu} \delta A_\mu.
\een
As is seen from this equation, $H$ does not depend on $\dot A_0$, which is general in the constrained systems, despite that eq.(3) cannot be solved to obtain $\dot A_0$.
This variation must be equal to direct variation \be
	\delta H = {\d H \over \d \pi^\mu} \delta \pi^\mu +  {\d H \over \d A_\mu} \delta A_\mu.
\ee
We must note that the variables ($\pi^\mu$ and $A_\mu$) cannot  be varied freely, but they must obey \[
	\delta\phi =  {\d\phi \over \d\pi^\mu} \delta \pi^\mu + {\d\phi \over \d A_\mu} \delta A_\mu = 0,
\]
which follows the constraint.  Therefore, using Lagrange's undetermined multiplier $\lambda$, we have \ben
	\dot A_\mu - \lambda {\d\phi \over \d\pi^\mu} = {\d H \over \d\pi^\mu},			\\
	- {\d L \over \d A_\mu} - \lambda {\d \phi \over \d A_\mu} = {\d H \over \d A_\mu}.
\een
By using Euler-Lagrange equation of motion, we obtain
\bea
	\dot A_\mu &=& {\d H \over \d \pi^\mu} + \lambda {\d \phi \over \d \pi^\mu},			\nonumber\\
	\dot\pi^\mu &=& -{\d H \over \d A_\mu} - \lambda {\d \phi \over \d A_\mu}.			\label{HJeq}
\eea
Now, this Hamilton-Jacobi's equation of motion is modified because of the constraint.  We can incorporate this modification by modifying the Hamiltonian to \[
	H \rightarrow H + \lambda \phi.
\]
Finally, the Hamiltonian may be written as \bea
	H = \int d^3x \Big[ {1\over2} \vec\pi^2 +{1\over 2} (\nabla \times \vec A)^2+ A_0 \dpi - A_0 \rho +\lambda\pi_0
		 + e  \inp j A \Big].			\label{15}
\eea
We will denote charge density $e j_0$ by $\rho$ sometimes.

Then, we define the Poisson brackets. \be
	\{A_\mu(\vec x), \pi^\nu(\vec y) \} = \delta_\mu^\nu \delta^3(\vec x - \vec y).			\label{4}
\ee
We must not use constraint(s) $\phi = 0$ in calculating Poisson brackets.  They are defined assuming that all the variables are independent.

Next, the constraint condition $\phi = 0$ must hold all over the time.  Therefore $ \dot\phi = 0, \ddot\phi = 0, \cdots$ must hold.  These conditions are called ``consistency conditions".  For here, we have \be
	0 = \dot{\pi}_0 = \{\pi_0, H\} = - (\dpi -\rho).			\label{12}
\ee
This is a new constraint.  The constraint $\phi = 0$ that was produced in going to canonical variables is called ``primary constraint", which can be plural, and the constraints produced from consistency condition like eq.(\ref{12}) are called ``secondary constraints".
Further, we must check if consistency conditions for secondary constraints hold, and have in our case \[
	{d\over dt}(\dpi -\rho) = e \inp\nabla j -\dot{\rho},
\]
where we have used one of the equations of motion $\dot{\vec\pi} = \vec\nabla\times\vec\nabla\times\vec A + e \vec j $.
This condition holds since we can expect $\d_\mu j^\mu = 0$ for electric charge.  Hence, we do not have a tertiary constraint.  Dirac conjectured that the consistency conditions end at a finite step likewise.

The undetermined multiplier $\lambda$ may be determined by a secondary constraint for some of the systems other than QED.
In the systems with gauge symmetry, $\lambda$ won't be determined by the consistency conditions, and $\lambda$ will be arbitrarily depend on time $t$ and space $\vec x$.

The Hamiltonian above, the Poisson brackets and the constraints will consistently describe the dynamics.

\subsection{ Quantization of constrained systems }

To quantize a system with gauge symmetry, people usually carried out gauge fixing and eliminated a part of variables.  Fixing gauge means adding as much conditions as necessary to fix the gauge freedom.

In this paper, we propose a way of quantization that does not fix the gauge.  We will simply replace\footnote{
Actually, Dirac seems to have thought of requiring $\phi|*\> = 0$ for the first class constraint $\phi$, not fixing the gauge(see p34, of \cite{Dir64}).
If one pursues that direction, we should already have reached this paper's conclusion that there exists physical freedom of electrostatic field and the conventional method of taking only the transverse modes is not correct.
It seems, however, that Dirac was not pursuing that direction. The author also could not find any other author who is pursuing.
Please do not confuse this method of imposing constraints with well-known Gupta-Bleuler's method, since it has a fault.
Problems with these conventional methods will be fully discussed in section 3.
}
the Poisson brackets eq.(\ref{4}) by \[
	[A_\mu(\vec x), \pi^\nu(\vec y)] = i\hbar \delta_\mu^\nu \delta^3(\vec x - \vec y),
\]
where we took Schr\"odinger picture.
A space is determined by this algebra, and a part of this space spanned by the states satisfying the following conditions will be defined as physical space.
\bea
	\pi_0|*\> = 0,			\nonumber\\
	(\dpi -  \rho)|*\> = 0.			\label{constraints}
\eea
If the constraints are satisfied at a certain time, it will be satisfied all the time because the commutators between the Hamiltonian and the constraints will become linear combinations of the constraints owing to the consistency conditions.

A comment should be made on imposing constraints.  For example, let us suppose that we have $\pi_0|*\> = 0$, then we will have $ \<*|(\pi_0 A_0 - A_0 \pi_0)|*\> = 0$ and it seems to be inconsistent with $[A_0, \pi_0] = i\hbar$.
We can see that this is actually not a problem by considering an example in elementary quantum mechanics.
Let us consider an ordinary free particle: \[
	H =  p^2/2,
\]
where \[
	[x, p] = i\hbar.
\]
In this theory, the vacuum state specified by \[
	p |0\> = 0
\]
is simply accepted, and this state is obviously analogous to the constraint $\pi_0|*\> = 0$.  Then the wave function of this state is
\be
	\psi(x) = 1,			\label{5}
\ee
which is infinitely and flatly extending, and observed value of  $ x $ will be extending infinitely.  Therefore $ \<0|[p, x]|0\> = i\hbar $ can hold as a product of zero and infinity.
Strictly speaking, the state like eq.(\ref{5}) is illegal as a quantum state because it is not square integrable.  But we have no problem as far as we don't forget that this is an approximation for a spatially extended state.\footnote{
This understanding does not hold if the constraint is second class (see section 3).  The constraints here are first class.
}
Another way to do with the problem is introducing a fictitious infinitesimally small mass $\omega$ and letting \[
	H = {1\over2} p^2 + {1\over2} \omega^2 x^2.
\]
Its ground state approximates the state of $p = 0$ because \ben
	x &=& (a + a^\dagger)/\sqrt{2\omega },			\\
	p &=& -i\sqrt{\omega/2}(a - a^\dagger),
\een \[
	[a, a^\dagger] = 1,
\]
and the ground state is defined by \[
	a|0\> = 0,
\]
and then we have \[
	p|0\> =  -i \sqrt{\omega/2} (a - a^\dagger)|0\> \approx 0.
\]
This approximation holds if the relevant time scale is smaller than $\omega^{-1}$.
The state that corresponds to $\tilde\psi(p) = \delta(p-k)$ may be further constructed as \[
	e^{ikx}|0\>.
\]
This state is approximately an eigenstate of $H = p^2/2$ and will be kept unchanged if $t \ll \omega^{-1}$.  This state is exactly an eigenstate when $\omega \rightarrow 0$.

We may think that $\pi_0|*\> = 0$ is fulfilled likewise.
Then the wave function for $A_0$ is extending uniformly from $-\infty$ to $\infty$, which may be interpreted that all the gauge equivalent configuration is uniformly summed.
(Uniform summation of gauges was an ideal of path integral, which could not be realized.)

The above completes definition of quantization.  The calculation in the following subsections follows this definition.

\subsection{ Free QED}

Here, we Fourier decompose the fields as \bea
	\psi(\vec x) &=& \int {d^3p \over \sqrt{(2\pi)^3 2E_p}} \sum_s (b_{p, s} u(\vec p, s) \exp(i\inp p x)+ d_{p, s}^\dagger v(\vec p, s) \exp(-i\inp p x)),			\nonumber\\
	A_i(\vec x) &=&  \int {d^3k \over (2\pi)^{3/2}} \Big[ \sum_{\sigma=+,-} {1\over \sqrt{2|\vec k|}} \exp(i\inp k x) \e_i(\vec k,\sigma) a_\sigma(\vec k)		\nonumber\\
		&& + {1\over \sqrt{2\omega }} \exp(i\inp k x) \e_i(\vec k, L) a_L(\vec k) + {\rm c.c.} \Big],			\label{A}		\\
	\pi_i(\vec x) &=& \int {d^3k \over (2\pi)^{3/2}} (-i) \Big[\sum_{\sigma=+,-} {\sqrt{|\vec k|}\over \sqrt 2} \exp(i\inp k x) \e_i(\vec k,\sigma) a_\sigma(\vec k)			\nonumber\\
		&&	+ {\sqrt \omega\over \sqrt 2} \exp(i\inp k x) \e_i(\vec k, L) a_L(\vec k) - {\rm c.c.} \Big],							\nonumber
\eea
where \ben
	&& \vec\e(\vec k,\sigma) \cdot \vec\e(\vec k,\sigma') = \delta_{\sigma\sigma'},		\\
	&& \e_0(\vec k,\sigma) = 0,			\\
	&& \vec\e(\vec k, L) = \vec k/|\vec k|,
\een
and $\sigma$ takes one of $+,-,$ or $L$.  From $[A_i, \pi_j] = i\hbar \delta_{ij}$, we have \[
	[ a_\sigma(\vec k), a_{\sigma'}^\dagger (\vec k)] = \delta_{\sigma\sigma'} \delta^3(\vec k - \vec k').
\]
Later, the vibration mode created by $ a^\dagger_+, a^\dagger_-$ will be called $T$ mode or transverse mode since it is transverse to $\vec k$, and the mode created by $a^\dagger_L$ will be called L mode or $\vec k$-proportional mode.

The Hamiltonian for the free field is obtained by letting $e = 0$ into eq.(\ref{15}).
Further, the $A_0(\dpi - \rho)$ term in $H$ does not have any effect, since every physical state should satisfy the constraint $(\dpi - \rho)|*\> = 0$.
(Since $\pi_0 = 0$, the value of $A_0$ is uniformly and infinitely distributed.  We don't, however, have any problem with this non-zero $A_0$ owing to this constraint.)  Then we have a free Hamiltonian \ben
	H_0 &=& \int d^3x \Big[ {1\over 2} \vec\pi^2 + {1\over 2} (\vec \nabla \times \vec A)^2 + {1\over 2} \omega^2 \vec A^2 \Big ]			\\
	&=& \int d^3p \ \Big[\sum_{\sigma=+,-} |\vec p| a_\sigma^\dagger(\vec p) a_\sigma(\vec p)	+ \omega a_L^\dagger(\vec p) a_L(\vec p) \Big],
\een
where we have added a fictitious small mass $\omega$.

We define the vacuum $|0\>$ for $H_0$ by \be
	\begin{array}{l}
		\pi_0|0\> = 0,			\\
		a_+|0\> = 0,			\\
		a_-|0\> = 0,			\\
		a_L|0\> = 0.
	\end{array}			\label{vac}
\ee
The third condition $a_L|0\> = 0$, which is easier for some of calculation, is an approximation for
\[
	\dpi|0\> \propto (a_L^\dagger(\vec p) - a_L(\vec p))|0\>  = 0,
		\hspace{1cm} (\forall \vec p).
\]
This approximation is justified because the difference between the ground state $a|0\> = 0$ of a harmonic oscillator with a very small frequency $\omega$ and the state of zero canonical momentum $|\pi=0\> $ is very small, as mentioned in the previous subsection.
Further, the $\lambda\pi_0$ term in $H$ does not have any effect owing to the constraint $\pi_0|0\> = 0$.

\subsection{ Electrostatic field
--- an overlooked physical degree of freedom in quantum theory}

Let us consider the quantum state of electrostatic field for a system with a classical and localized charge distribution $\rho(\vec x)$.
Or we may think that we are observing an electron from distant enough point.

Now, the $\vec k$ proportional part of $\vec E$ should satisfy the constraint $(i \inp k E(\vec k) + \rho(\vec k))|*\> = 0$.
Such a state may be constructed as \be
	|\psi\> = \exp \Big\{\sum_k \rho(\vec k) \frac{\inp k A}{\vec k^2} \Big\}|0\>.			\label{1}
\ee
This holds exactly, but it becomes an approximation if the fictitious mass $\omega \neq 0$ and it holds only when $t \ll 1/\omega$.
(In the limit of $\omega\rightarrow 0$, the probability of appearance of L mode photon get very large and infinitely many number of them will appear.  It naturally occurs when a continuous observable is mimicked by a harmonic oscillator.)
Please note that this state is already normalized.  That is because commuting that exponential of $(a_L^\dagger + a_L)$ in eq.(\ref{1}) to have only normal ordered terms gives a normalization factor like $\exp(-\rho^2/\omega)$.
This state is already normalized even though higher order multiparticle states have larger factor in the expansion.  This is because commuting the terms in the formula (\ref{1})
so as to have only normal ordered terms gives overall factor $\exp(-\rho^2/\omega)$, which is appropriate for normalization.

In this formalism, the electrostatic field $\vec E(x) = \vec\pi(x)$ is plainly a local operator and a observable\footnote{
Contrarily, in Conventional quantum field theory (QFT), the electrostatic field was not presented as an observable.  
In addition, conventional QFT lacked the notion of observable even if not completely but mostly and is only good at scattering calculation.
}.
We note that the $\vec k$ proportional part of electric field $\vec E$ takes an continuous value even after quantization as well as $p$ in $H=p^2/2$.

Let us suppose that an observation was made on this state.
We multiply the operator $\vec E = \vec \pi$ to state $|\psi \>$.  We have \[
	\sum_{k} \exp(-i\inp k x) \Big\{ i \rho(\vec k) {\vec k \over {\vec k}^2} \Big\}|\psi\>,
\]
using the commutation relation $[A_i, \pi_j] = - i\hbar\delta_{ij}$.  We see that $|\psi\>$ is an eigenstate of $\vec E$ and that the strength $|\vec E|$ is proportional to the inverse square of the distance from a far enough charge.

Contrarily, in the conventional quantum theory of gauge field, a classical state with an electric field is not a state in the quantum sense, and electric field $\vec E(\vec x)$ is not a quantum observable nor a local operator.
For instance, let us take Coulomb gauge $\inp \nabla A = 0$, where $\vec k$ proportional part of $\vec A$ is removed.
Then using the constraint (\ref{12}), we have $\dpi = \vec\nabla \cdot(-\dot{\vec A} - \vec\nabla A_0) = - {\vec \nabla}^2 A_0 = \rho$.
Solving this, we obtain \[
	A_0(\vec r) = \int {d^3\vec{r'}} \rho(\vec{r'}) {1\over4\pi|\vec r - \vec{r'}|}.
\]
Substituting this to $A_0$ dependent part of the Hamiltonian: $(\vec\nabla A_0)^2/2 + (\dpi-\rho )A_0$, we obtain \be
	H_{\rm Coulomb} = {1\over2}\int d^3\vec x \int d^3\vec x' \rho(\vec x) \rho(\vec x') {1\over4\pi|\vec x - \vec x'|}.			\label{HCoulomb}
\ee
This part of $H$ gives the effect of electrostatic field, but now the classical state variable of electrostatic field is lost, and electrostatic force is now a superluminal infinite distant force between the electrons.
The conventional Coulomb gauge prescription is quantizing this classical Hamiltonian from which the electrostatic field is removed.  Then the electrostatic field is not an operator and does not have corresponding quantum state.
Since other well-known gauge prescriptions are the same in that they leave only transverse mode, they have the same problem of not having electrostatic field.

\subsection{ Gauge Transformation }

In classical mechanics, the time variation of $A_0$ depends on $\lambda$ as $\dot A_0 = \lambda$, which is a part of the canonical equations of motion (\ref{HJeq}).
Further, $\vec A$ depends on $\lambda$ through $A_0$ in $\dot{\vec A} = - \vec\pi - \vec\nabla A_0$, which is one of equations of motion too.
If one change $\lambda$ by $\Delta\lambda= \ddot \theta$, the change of $A_0$ is
$\Delta A_0 = \dot\theta$ and
$\Delta\dot{\vec A} = - \vec\nabla \Delta A_0$.
Then, we have $\Delta A_\mu = \d_\mu \theta$, which is the familiar gauge transformation.

In the Schr\"odinger picture of quantum theory, we can't apply any time varying gauge transformation since the operators do not vary in time.
A gauge transformation by time independent $\theta(\vec x)$ can be performed by applying a generator $\int d^3\vec x\; \theta(\vec x)\phi(\vec x),$ $(\phi = \dpi -\rho)$ and we get \ben
	\delta \vec A &=& [\phi\theta , \vec A] = \vec\nabla \theta		\\
	\delta\psi &=& i\theta\psi.
\een
We don't have transformation for $A_0$ here, but we don't have problem because we don't need $A_0$ in constructing and time-developing physical states as we will see later.
This transformation does not change physical state since $\phi|*\> = 0$, and it won't change any observed value of physical observable that is gauge independent.

\delete{
In Heisenberg picture, if we apply \[
	\int d^3x (\d_0\theta  \pi_0 + \theta(\vec x,t)\phi(\vec x,t))
\]
as the generator for time varying $\theta(\vec x, t)$, we have \ben
	\delta\vec A &=& [\phi\theta ,\vec  A] = \vec\nabla \theta,			\\
	\delta A_0 &=& \d_0\theta.
\een
Any physical state is invariant under this generator, since the constraints $\pi_0, \phi$ does not change by time even in Heisenberg picture.
Any gauge-independent combination of the Heisenberg operators does not be affected by this gauge transformation too.
} 

\subsection{ The initial and the final states }



Now we would like to start discussing interaction.
For that purpose, we must, in advance, clarify what is the ``states", which appear before and after interaction.
We must know what ``a particle" is in field theory, because the main actor in  scattering and the other physical processes is the ``particles".
In the conventional quantum field theory, answering this question and describing quantum states of a particle was vaguely evaded.  Instead, people used  LSZ's reduction formula.

The advantage in departing from LSZ formalism is that
finite time transient phenomena can be now predicted by this formalism, which contrasts to that only the scattering processes between the infinite past and the infinite future may be calculated in LSZ formalism.

We will begin by perturbatively constructing the vacuum state, 1-particle states, and multiparticle states in order.

\subsubsection{ The Vacuum}

We defined the vacuum as the lowest energy eigenstate of $H$ satisfying the constraints (\ref{constraints}).
Perturbatively, the solution is \[
	|0\> - {1\over H_0 - i\epsilon} V|0\>
	+ {1\over H_0- i\epsilon} V{1\over H_0 - i\epsilon} V|0\> + \cdots,
\]
where $|0\>$ is the free field vacuum defined by eqs.(\ref{vac}), and interaction part in $H$ is $V= e\int d^3x\; \inp j A$ since we may ignore the constraint-proportional part of the Hamiltonian (\ref{15}).
(The counterterms for energy renormalization in the 2nd or higher orders are not shown.)
This expression may as well be written as \[
	e^{-i\infty (H - i\epsilon)}|0\>.
\]

\subsubsection{One-Particle States}

An one-particle state may be treated as an eigenstate in the same way as in ordinary quantum mechanics.
Perturbatively, the eigenstate is \bea
	|\vec k\> &=& b_k^\dagger|0\> - \invHE V b_k^\dagger|0\>		\nonumber\\
		&& + \invHE V\invHE V b_k^\dagger|0\> + \cdots.	\label{9}
\eea
This 1-particle state may be also written as $e^{-i\infty (H - i\epsilon)} b_k^\dagger|0\>$.

This eigenstate thus made satisfies the constraint $(\dpi - \rho)|*\> = 0$.  Let us prove it up to the first order for example.  The following two term will appear by applying  $(\dpi - \rho)$ to the perturbative state (\ref{9}).  One is \[
	-\rho_k b_p^\dagger|0\> = e b_{p'}^\dagger{u(\vec{p'})\over \sqrt{2E_{p'}}}\gamma _0 {\bar u(\vec p)\over E_p}|0\>.
\]
The other is \[
	-e \dpi_k \times{1\over p'_0 + \omega - p_0} b_{p'}^\dagger{u(\vec{p'})\over \sqrt{2E_{p'}}} {\inp A \gamma} {\bar u(\vec p)\over E_p}|0\>,
\]
where $E_p \equiv \sqrt{\vec p^2 + m^2}$.
The latter term cancels the former term by commuting
$\dpi_k$ onto right and using $\dpi_k|0\> = 0$ and $k_i u(\vec {p'})\gamma _i \bar u(\vec p) = (E_{p'} - E_p) u(\vec {p'})\gamma _0 \bar u(\vec p)$.
This constraint ensures that electrostatic field exists.  The other constraint $\pi_0$ holds trivially since eq.(\ref{9}) does not have $A_0$ in it.

Then let us check if there actually exists the electrostatic field in the first order.  By explicitly writing the term in eq.(\ref{9}), we have \[
	{1 \over E_{p-k} + \omega - E_p } \int d^3x \; e \bar\psi \gamma_i A_i \psi(\vec x) \; b_p^\dagger|0\>.
\]
Here we will take $ A_L \equiv \vec \e(\vec k, L) \cdot \vec A$ part of  $A_i$ and will use $\bar u(p-k) k_i \gamma_i u(p) = (E_p - E_{p-k}) \bar u(p-k) \gamma_0 u(p)$ to cancel $ 1 / (E_{p-k} + \omega - E ) $.
\be
	=   \sum_k {1\over |\vec k|} \rho(-\vec k) (\inp{\e(\vec k, L)}{A(\vec k)}) \; b_p^\dagger|0\>. 			\label{521}
\ee
This formula agrees with the first order of the state (\ref{1}), where the electric charge density $\rho$ here is $ - e \bar\psi \gamma_0 \psi$.

The 1-photon state may as well be written as \ben
	&& a_\sigma^\dagger(\vec k)|0\> - \invHE V a_\sigma^\dagger(\vec k)|0\>	\\
	&& + \invHE V\invHE V a_\sigma^\dagger(\vec k)|0\> + \cdots,
\een
where $\sigma$ must be $+$, or $-$, i.e.\ only transverse modes.  That is because L mode does not propagate the space by itself.  Charge $\rho(\vec k)$ must exist for the L mode photon to exist.

\subsubsection{ Multiparticle States}

We will obtain the multiparticle states here, which is necessary because the initial and the final states are multiparticle states in scattering.
Multiparticle states may be constructed by placing localized particles that are separated spatially.
At first, a localized 1-particle wave packet with its wavefunction $\varphi(\vec k)$ may be constructed by superposing states of different momentum. \[
	|\varphi\> = \sum_{\vec k} \varphi(\vec k) \tilde b^\dagger(\vec k) |0\>,
\]
where $(1- (H_0-E )^{-1} V + \cdots ) b_k^\dagger$ was ordered so as to have only normal ordered terms using the commutation relations, and then the part that does not include annihilation operators $a_k, b_k$ is defined as $\tilde b^\dagger(\vec k)$.  Then we may construct a 2-particle state \[
	|\varphi_1, \varphi_2\> = \sum_{\vec k_1, \vec k_2} \varphi_1(\vec k_1) \varphi_2(\vec k_2) \tilde b^\dagger(\vec k_1) \tilde b^\dagger(\vec k_2) |0\>
\]
with separated and spatially non-overlapping wave packets.
Any scattering amplitude may be obtained evolving this state in time.

Then we will mainly discuss in the plane wave states: \[
	|\vec k_1, \vec k_2\> \equiv \tilde b^\dagger(\vec k_1) \tilde b^\dagger(\vec k_2) |0\>
\]
since it is easier to calculate.  Interaction between the particles before $t=0$ may be neglected if the space is infinitely large.

\subsection{ Scattering }

Here let us calculate probability of Compton scattering for example.  The initial state is a state with one electron and one photon $|\vec p; \vec k\>$.  After a long time $T$, we may obtain the amplitude by decomposing \[
		e^{-iHT}|\vec p; \vec k\>
\]
into multiparticle states $|\vec k_1,\vec k_2,\cdots\>$.
In other words, we should calculate \be
		\< \vec k_1,\vec k_2,\cdots|e^{-iHT}|\vec p; \vec k\> \equiv {\cal M'},			\label{14}
\ee
where $|\vec k_1,\vec k_2,\cdots\>$ is a multiparticle state defined in the previous subsection.
People sometimes call such transition amplitude ``S-matrix" regarding this to be a matrix.

Then we calculate \ben
	\lefteqn{e^{-iHT}|\vec p, \vec k\>} 			\\
	&=& \Big\{1+ \cdots + (-i)^2\int ^T_0 dt_2\int dt_1 e^{-iH_0(T-t_2)}
	V e^{-iH_0(t_2-t_1)} V e^{-iH_0 t_1} \Big\}|\vec p, \vec k\>
\een
up to the second order in $e$.
Here we don't need the constraint-proportional part in $H$ such as $A_0(\dpi-\rho )$.  This is because $(\dpi-\rho )|{\rm phys}\> = 0$ at the initial time since we use the initial states constructed in the previous subsection so as to fulfill the constraints.  Then the constraints hold all the time because $H$ and constraints commute owing to the consistency conditions.  Therefore the constraint-proportional part in $H$ does not have any effect in calculating infinitesimal time evolution.
(We may comment that the calculation proceeds only with gauge independent states in our formalism.)
Then we are left with only the term \[
	V= e \int d^3x \ \inp j A
\]
as the interaction vertex.

The 1-electron state in the initial state includes virtual photons perturbatively.  This is different from conventional field theory.  After short time evolution, i.e.\ in a transient phenomena, this virtual photon may be observed in the final state.  On the other hand, when time $T$ is large enough, scattering probability is proportional to $T$.  Then we may ignore the effect of this virtual photon because its effect is only constant in $T$.  Then in $T\rightarrow \infty$, we finally have the same result as the conventional perturbation, in which the initial and final states are free fields.
Therefore we have \bea
	{\cal M'}  \approx  \<0| b_{p'} a_{\sigma'}(\vec k') (-ie)^2 e^{-iH_0 T} \int d^4x\; \bar\psi_I \inp{A_I}{\gamma} \psi_I(x)			\nonumber\\
		\times \int_{x_0>y_0} d^4y\; \bar\psi_I \inp{A_I}\gamma \psi_I(y)\; b^\dagger_{p} a_\sigma^\dagger(\vec k)|0\>,			\label{43}
\eea
where $ I $ stands for interaction picture, for example $A_{Ij}(t) = e^{i H_0 t} A_j e^{-i H_0 t}$.
The $e^{-iH_0T}$ factor in this expression may be ignored since it merely gives a phase factor.

For eq.(\ref{43}), there is two ways of how to contract $b_{p'}, b^\dagger_{p}$ and the two vertices.
Using time ordered product, this two terms may be written into one formula: \ben
	{1\over 2\sqrt{E_{p'} E_p}} \bar u(\vec {p'}, s')
	\<0| a_{\sigma'}(\vec k') (-ie)^2 {\rm T} \int d^4x\; e^{+i E_{p'} x_0} \exp(-i\inp{{p'}}x) \inp{A_I}{\gamma} \psi_I(x)			\\
	\times \int d^4y\; \bar\psi_I(y) \inp{A_I}\gamma e^{-i E_p y_0}\exp(i\inp {p} y)  a_\sigma^\dagger(\vec k)|0\> u(\vec p,s),
\een
where ${\rm T}$ signifies time-ordered product.
Note that restriction $x_0 > y_0$ in the integration is removed now and $x_0$ and $y_0$ run from $-\infty$ to $-\infty$.  Considering two possible contraction of the photons, it equals to \bea
	{1\over 4\sqrt{E_{p'} E_p \omega_{k'} \omega_k}}
	\int d^4x \int d^4y\; e^{+i E_{p'} x_0 - i E_p y_0}
e^{-i \inp {{p'\!}} x + i\inp p y}			\nonumber\\
	\times \{ \e_j(\vec k', \sigma') \e_i(\vec k,\sigma) e^{+i \omega_{k'} x_0 -i \omega_k y_0} e^{-i\inp {k'} x + i\inp k y}			\nonumber\\
		+ \e_i(\vec k', \sigma') \e_j(\vec k,\sigma) e^{+i \omega_{k'} y_0 - i \omega_k x_0} e^{+i\inp {k} x - i\inp {k'} y} \}			\nonumber\\
	 \times (-ie)^2 \bar u(\vec {p'}, s')\<0|\gamma_j {\rm T} \psi_I(x) \bar\psi^I(y) \gamma_i |0\> u(\vec p,s),
	\label{624}
\eea
where \[
	\<0|{\rm T}\psi_I(x) \bar\psi^I(y) |0\> \equiv S_F(x-y)
\]
is called ``Feynman propagator" for electron, and $\omega_k \equiv \sqrt{\vec k^2 + \omega^2}$ for $\sigma \ne L$ and $\omega_k = \omega$ for $\sigma = L$.

By changing the integration variables as $y \rightarrow y+x$,
\ben
	 {1\over 4\sqrt{E_{p'} E_p \omega_{k'} \omega_k} }
	\int d^4x \int d^4y\;  e^{-i(E_p-E_{p'})x_0 - iE_p y_0} 
e^{-i (\vec {p'} - \vec p)\cdot \vec x + i\inp p y}			\\
	\times \{ \e_j(\vec k', \sigma') \e_i(\vec k,\sigma) e^{-i (-\omega_{k'}+\omega_k)x_0 -i \omega_k y_0} e^{-i (\vec k' - \vec k)\cdot \vec x + i\inp k y}				\\
		+ \e_i(\vec k', \sigma') \e_j(\vec k,\sigma) e^{+i \omega_{k'} y_0 -i (\omega_k - \omega_{k'}) x_0} 
e^{+i (\vec k - \vec k')\cdot \vec x - i\inp {k'} y} \}			\\
	  \times (-ie)^2 \bar u(\vec {p'}, s') \gamma_j S_F(-y) \gamma_i u(\vec p,s).
\een
By integrating $x_\mu$, we get the factor $(2\pi)^4 \delta^4(p'+k' -p -k) $, and $y_\mu$ may be integrated separately to get
\bea
	\lefteqn{(-ie)^2 {1\over 4\sqrt{E_{p'} E_p \omega_{k'} \omega_k} }  (2\pi)^4 \delta^4(p' + k' - p - k)}			\nonumber\\
		&&\times \bar u(\vec {p'}) \Big\{ \gamma _\nu{ i\over \p + k\ - m }\gamma _\mu 
		+ \gamma _\mu{ i\over \p - \k' - m }\gamma _\nu \Big\} u(\vec p) \e^*_\nu(\vec k', \sigma') \e_\mu(\vec k, \sigma) ,
\eea
where $\vec q = \vec p - \vec k $.\footnote{
The textbooks to date derived the same expression applying LSZ reduction formula\cite{LSZ} to the n-point Green function, and the asymptotic condition (or adiabatic hypothesis) was required in the derivation of the reduction formula.  We don't need that hypothesis.  From our viewpoint, their hypothesis about the asymptotic states was needed because we did not know what the initial and the final states are.
%
}
Note that we used a little trick here, which was commonly used in the textbooks.  We should have paid more attention to that change of integration variable, but we were cheated because the integration ranged from $-\infty$ to $\infty$.
Actually, the two integration cannot be preformed independently since the upper bound of $y'$ integration was $y'+x < T$.  We needed a correction when $x_0$ is close to  final time.
This correction, however, is small compared to very large $T$ as well as that correction from perturbative correction for the initial state, which was ignored.  This approximation can be also mentioned as ``taking the leading term in the limit of $T \rightarrow 0$".

The scattering amplitude obtained above have no L mode excitation in the final state.  (Please note that the approximation taken above is intentionally introduced so as to reproduce the conventional result for the purpose of comparison.) \ben
	{\cal M}(\sigma'=L) \simeq 
-i \omega\; \bar u(\vec {p'}) \Big\{ \gamma_0{1\over  \p + \k-m }\gamma_\mu			
		+ \gamma_\mu {1\over  \p - \k' - m } \gamma_0 \Big\} u(\vec p) ,
\een
where we have used $p_0 = E_p, k_0 = \omega_k, k'_0 = \omega$, and $\mu \neq 0 $.  Here the zeroth components of $k_0, k'_0$, etc.\ are $|\vec k|$ for T polarization but $\omega$ for L polarization.  Further, we have used $ (E_p\gamma_0 -p_i \gamma_i - m)u(\vec p) = 0$ and so on.
Then the amplitude ${\cal M'} \rightarrow 0$ as $ \omega\rightarrow 0 $.

We will have L mode excitations, if we don't take the above approximation on the other hand, and it will produce the electrostatic field around the final state particle.
When the final state photon is L-polarized, we rewrite eq.(\ref{624}) now paying attention to the range of the integration as \ben
	 {(-ie)^2 \over 4\sqrt{E_{p'} E_p \omega_{k'} \omega_k}}
	\int_{T > x_0} d^4x \int_{x_0 > y_0 > 0} d^4y\; e^{-i (E_{p'}+\omega) (T - x_0)}
	\bar u(\vec {p'}, s') \e_j(\vec k', L)			\\
	\times \sum_{s_1} \Big\{
		\gamma_j u(\vec p + \vec k, s_1) {1\over 2 E_{p+k}} e^{-i E_{p+k} (x_0 - y_0)} \bar u(\vec p + \vec k, s_1) \gamma_i			\\
		+ \gamma_i u(\vec {p'} - \vec k, s_1) {1\over 2 E_{p'-k}} e^{-i (E_{p'-k}+\omega+\omega_k) (x_0 - y_0)} \bar u(\vec {p'} - \vec k, s_1) \gamma_j
	\Big\}			\\
	\times \e_i(\vec k,\sigma) u(\vec p,s) e^{-i(E_p+\omega_k) y_0},
\een
where the intermediate positron was ignored because it is irrelevant in low energy.  Performing the time integration, it equals to \ben
	\lefteqn{ {(-ie)^2 \over 4\sqrt{E_{p'} E_p \omega_{k'} \omega_k}} {1\over E_{p'}+\omega - E}
	\bar u(\vec {p'}, s') \e_j(\vec k', L)}			\\
	&& \times \sum_{s_1} \Big\{
		\gamma_j u(\vec p + \vec k, s_1) {1\over 2 E_{p+k}} {1\over E_{p+k} -E} \bar u(\vec p + \vec k, s_1) \gamma_i			\\
	&&	+ \gamma_i u(\vec {p'} - \vec k, s_1) {1\over 2 E_{p'-k}} {1\over E_{p'-k}+\omega+\omega_k - E} \bar u(\vec {p'} - \vec k, s_1) \gamma_j
	\Big\}			\\
	&& \times \e_i(\vec k,\sigma) u(\vec p,s),
\een
where $E \equiv E_p + \omega_k$.
Using $\e_i(\vec k, L) = k_i/|\vec k|$, $\gamma_i p_i u(\vec p) = (\gamma_0 p_0 -m)u(\vec p) $, etc., we obtain \ben
	 {(-ie)^2 \over 4\sqrt{E_{p'} E_p \omega_{k'} \omega_k} |\vec k|}
	\bar u(\vec {p'}, s') \gamma_0
	\sum_{s_1}  {u(\vec p + \vec k, s_1) \bar u(\vec p + \vec k, s_1)\over 2 E_{p+k} (E_{p+k} -E)}  \gamma_i
	\e_i(\vec k,\sigma) u(\vec p,s).
\een
This is a transition matrix element.  The state that will give this matrix element is \bea
	|\psi_{\rm f}\> &=&   \sum_k {1\over |\vec k|} \rho(-\vec k) ({\vec \e(k, L)}\cdot {\vec A(\vec k)})|\psi_{\rm m} \>,
\eea
where \bea
	\rho(\vec k) &=& \frac{-e}{2\sqrt{E_{p} E_{p+k}}} \sum_{p, s',s_1} \bar u(\vec {p}, s') b_{p,s'}^\dagger \gamma_0 u(\vec p + \vec k, s_1) b_{p, s_1},		\nonumber\\
	|\psi_{\rm m}\> &=& {e \over 2\sqrt{\omega_k E_{p+k}}} \sum_{s_1} b_{p+k, s_1}^\dagger |0\> {1\over E_{p+k} -E} \bar u(\vec p + \vec k, s_1) \gamma_i
	\e_i(\vec k,\sigma) u(\vec p,s) ,			\nonumber
\eea
and we may replace $\rho$ using $\rho(\vec x) = -e \bar\psi(\vec x) \gamma_0 \psi(\vec x) $ in the first equation.
Thus this L mode is a part of the 1-electron state at the final time.
This L mode part agrees with the electrostatic part of the first order 1-electron state eq.(\ref{521}).
Such a contribution in that the pole of the electron propagator is cancelled means that the state is concentrated at $t=T$ in terms of the $\int dx_0$ integral.
In this four-point case, this final state is the one for the process $e+\gamma \rightarrow e$, and then this process will not occur because of conservation of energy and momentum.  This contribution is actually physical in 5-point case or more.
\label{QED-L}

Other physical processes such as $e^+ e^-$ pair annihilation may be computed in the same way.

\subsection{ Feynman rules }

The calculation in the previous subsection may be generalized to the higher order using Wick's theorem.
Resulting enumeration rule of the contraction is called ``Feynman rules".  Let us write down it here.

Using the expansion formula (\ref{A}), here we obtain \ben
	&& \< 0| {\rm T} A_{Ii}(x)  A_{Ij}(y) |0\> \equiv \Delta_{ij}(x-y) 		\\
	&& = \int {d^3p\over (2\pi)^3 }
		\Big[ {1\over 2|\vec p|} P_{ij}(\vec p)  \exp(+i|\vec p|(x_0-y_0) - i\vec p \cdot (\vec x - \vec y))\theta(x_0 - y_0)			\\
	&&	+ {1\over 2\omega } {p_i p_j\over|\vec p|^2}  \exp(+i \omega(x_0-y_0)-i\vec p \cdot(\vec x - \vec y))\theta(x_0 - y_0)
		+ ({\rm c.c.})\times\theta(y_0 - x_0) \Big],			\label{82}
\een
which is possible for L mode owing to the fictitious mass $\omega$, and where \ben
	&& P_{ij}(\vec p) = \sum_{\sigma = \pm} \e_i(\vec p,\sigma) \e_j(\vec p,\sigma)^* = \delta_{ij} - {p_i p_j\over|\vec p|^2}, 		\\
	&& P_{0i}(\vec p) = P_{i0}(\vec p) = P_{00}(\vec p) = 0.
\een
Then we have \bea
	\Delta_{\mu\nu}(x-y) = \int {d^4q\over (2\pi)^4} {i\over q^2-\omega^2 + i\epsilon} \Big\{\delta_{ij}
		- { q_i q_j \over q_0^2-\omega^2 + i\epsilon} \Big\} e^{iq(x-y)}			\label{7}
\eea
where right-hand side is regarded to be zero if $\mu = 0$ or $\nu = 0$.
This is the propagator for photons.

Now the Feynman rules reads:	\hfil\break
for photon propagator, $\displaystyle {i\over q^2+i\epsilon} \Big( \delta_{ij} - \frac {q_i q_j} {q_0^2 -\omega^2 + i\epsilon} \Big) $;	\hfil\break
for electron propagator, $\displaystyle {i \over \p -m +i\epsilon}$;	\hfil\break
for vertex, $(-ie)\gamma _\mu$;	\hfil\break
for a electron loop, $-1$;	\hfil\break
and the factor for the external lines can be read from the result of the previous subsection.

Further, it is useful to rewrite the terms in the braces in eq.(\ref{7}) as \ben
	\delta_{ij} - {q_i q_j \over q_0^2-\omega^2 + i\epsilon}
		&=& - g_{\mu\nu} + { q_0 q_\mu n_\nu+ q_0 n_\mu q_\nu - q_\mu q_\nu\over q_0^2-\omega^2 + i\epsilon},
\een
and we may ignore $k_\mu$-proportional part of the propagator, because the source term $j_\mu$ for $A_\mu$ obeys $\d_\mu j^\mu = 0$ owing to Ward-Takahashi identity\cite{WT}.
Then, we have \[
	-i\Delta_{\mu\nu}(x-y) = \int {d^4q \over (2\pi)^4} {g_{\mu\nu} e^{iq(x-y)} \over q^2-\omega^2 + i\epsilon} ,
\]
which is equivalent to Feynman gauge propagator in the case of Abelian gauge theory.

\subsection{ Calculation of electrostatic energy }

In this subsection, we calculate Coulomb energy, or Coulomb force, between two electrons using perturbation.
Let us suppose that $|E\>$ is a state that consists of two fermions and approximately have energy $E$.
The following approximate expression suffices because we are content with the leading contribution here. \[
	|E\> = \sum_{\vec p_1, \vec p_2} \psi(\vec p_1) \exp(i \inp{p_2}{R}) \psi(\vec p_2) \;
	b^\dagger_{p_1} b^\dagger_{p_2}|0\>,
\]
where $E = E_{p_1}+E_{p_2}$ and $\psi$ is a wave function, which represents a wave packet of an electron.
We suppose that the dimension of the wave packet $L$ is much smaller than the separation $|\vec R|$ between the fermions.

According to the standard procedure of perturbation, energy shift of the 2nd order is \[
	E_2 \equiv \<E|V {1\over H_0-E+i\epsilon} V|E\>.
\]
Since the L mode corresponds to the electrostatic field, here we take only that part in $V$. \ben
	V &=& e \int d^3p\; d^3k\;
		 (a_L^\dagger(\vec k)+ a_L(\vec k)){1\over \sqrt{2\omega }}			\\
		&& \times b_{p-k}^\dagger {\bar u(\vec p-\vec k)\over \sqrt{2E_{p-k}}}  {k_i\over |\vec k|}\gamma _i  b_{p} {u(\vec p)\over \sqrt{2E_{ p}}} .
\een
Therefore, \ben
	E_2 &=& \<0|b_{p_1-k} b_{p_2+k} V {1\over H_0-E+i\epsilon} V b^\dagger_{p_1} b^\dagger_{p_2}|0\>			\\
		&=& {\bar u(\vec p_2+\vec k) k_i\gamma _i u(\vec p_2) \over 2\sqrt{E_{ p_2} E_{p_2+k}}}	
			 {e^2 \over 2\omega {\vec k}^2} \{ {1\over E_{p_1-k} + \omega -E_{p_1}} + {1\over E_{p_2+k} + \omega -E_{p_2} } \}			\\
			&& \times {\bar u(\vec p_1-\vec k) k_i\gamma _i u(\vec p_1) \over 2\sqrt{E_{p_1} E_{p_1-k}}} .
\een
Here we have \[
	E_{p_2 +k} + E_{p_1 -  k} = E_{ p_2} + E_{ p_1}
\]
as the energy of final and initial state is the same, and \[
	\bar u(\vec p_1- \vec k) k_i\gamma _i u(\vec p_1) = (E_{p_1-k} - E_{p_1}) \bar u(\vec p_1-\vec k) \gamma_0 u(\vec p_1).
\]
Using this, \[
	E_2 = e^2 {\bar u(\vec p_2 + \vec k) k_i\gamma _i u(\vec p_2) \over 2 \sqrt{E_{p_2} E_{p_2+k}}}
	 {1\over {\vec k}^2} 
	{\bar u(\vec p_1 - \vec k) k_i\gamma _i u(\vec p_1) \over 2\sqrt{E_{p_1} E_{p_1-k}}}
\]
as $\omega \rightarrow 0$.  When $\vec k^2$ is small enough, the expression above will be $\propto 1/\vec k^2$, and we obtain \ben
	E_2 &\propto& e^2 \int d^3p_1\; d^3p_2 \int d^3k\;
		\psi^*(\vec p_2 + \vec k) \psi^*(\vec p_1 - \vec k)
		\exp(-i\inp k R) \psi(\vec p_2) \psi(\vec p_1)			\\
	&\simeq& {e^2 \over 4\pi|\vec R|},
\een
which is the classical Coulomb force.

\delete{
\subsection{ Renormalization }
Our formalism needs renormalization of energy as well as conventional theory if we are to do loop correction.  For example, perturbative energy shift of 1-particle state \[
	\Delta E_2 = \<0|b_p V(H_0 - E -i\epsilon)^{-1} V b_p^\dagger|0\>
\]
diverges at the 2nd order.  We have to cancel this divergence to have finite physical mass by shifting the bare mass $m$ in the Lagrangian.  This manipulation is called ``renormalization".  We happened not to need it for photons because of symmetry.
On the other hand, We don't need ``wavefunction renormalization", even though we have divergence in the two-point function from the 2nd order (1-loop).  That is because quantum states can be normalized irrespective of renormalization in the Lagrangian.  We don't have to add counterterms to the Lagrangian or the Hamiltonian in calculating transition from the initial to the final state\footnote{
We don't need to do wavefunction renormalization, but this does not mean that we must not.  If one wants to compare the result with the conventional formalism, it may be easier to do with the renormalization.
}.
The photon field won't need wavefunction renormalization too.
Finally, interaction terms, such as 3-point vertex in QED, need counterterm to the Lagrangian as well as the conventional formalism.
} 

\subsection{ Summary }

In this section, we saw that the advantage of our formalism over the conventional one is that any configuration of electrostatic field may be understood as a quantum state and be written down.  Further, the state of gauge field theory at any time may be written down and its time evolution may be described.
	Owing to this, 1-electron states may be written down, and scattering amplitude can be calculated without LSZ reduction formula, which needs an unjustified hypothesis called ``asymptotic condition".  Accordingly, calculating transient process is possible since time interval does not have to be infinite, which was required in LSZ formula.
	Further, because we can write down the states now,
we can expect applying field theories to bound system, equilibrium system, and so on, at which the conventional field theory was not good.

\section{non-Abelian gauge theory}

The Lagrangian density for non-Abelian ($SU(N)$) gauge theories is \bea
	{\cal L} &=& -\frac{1}{4} {F_{\mu\nu}^a}^2 + g A_\mu^a j^{\mu a}			\nonumber\\
		&=& {1\over 2} (-\dot{\vec A^a} - \vec\nabla A_0^a + g f_{abc} \vec A^b A_0^c)^2		\\
		&& - {1\over 2} (\vec\nabla \times\vec A^a - {1\over 2} g f_{abc} \vec A^b\times\vec A^c)^2 + g A_\mu^a j^{\mu a},			\nonumber
\eea
where $a, b, c$ are indices of adjoint representation of the non-Abelian group, and \[
	F_{\mu\nu}^a = \d_\mu A_\nu - \d_\nu A_\mu + g f_{abc} A_\mu^b A_\nu^c,
\]
and $D_\mu$ is a covariant derivative: $D_\mu =\d_\mu- i g A_\mu$.  We are assuming fermions for the source term $A_\mu^a j^{\mu a}$.

The canonical variables are defined as \[
	\pi^{\mu a} = {\d L \over \d\dot A_\mu^a}.
\]
Then we have \ben
	\vec\pi^a &=& -\dot{\vec A} - \vec\nabla A_0 + g f_{abc} \vec A^b A_0^c,			\\
	\pi_0 &=& 0,
\een
and using this, we move to canonical formalism and obtain the Hamiltonian: \bea
	H = {1\over 2} (\vec\pi^a)^2 +{1\over 2} (\vec\nabla \times\vec A^a - {1\over 2} g f_{abc} \vec A^b\times\vec A^c)^2	+ \vec j^a \cdot \vec A^a 		\nonumber\\
		+ A_0^a ({\vec D\cdot\vec\pi}^a - \rho^a) +\lambda^a \pi_0^a,			\label{HnA}
\eea
where $\rho = g j_0$.
Quantization is defined as requiring commutation relation \[
	[ A_\mu^a(\vec x) ,\pi^{\nu b}(\vec y)]
	= i \hbar \delta_\mu^\nu \delta^{ab} \delta^3(\vec x - \vec y)
\]
and imposing constraints
\bea
	\pi_0^a |{\rm phys}\> &=& 0,			\nonumber\\
	(\vec D\cdot\vec\pi^a -\rho^a)|{\rm phys}\> &=& 0,		\label{11}
\eea
which is not different from the Abelian case.  This is the nonperturbative definition of quantization.

The vacuum $|0\>$ is defined by \ben
	\pi_0^a|0\> = 0,			\\
	(\dpi^a -  \rho^a)|0\> = 0,			\\
	a_+|0\> = 0,			\\
	a_-|0\> = 0.
\een
for free field ($g=0$).
The quantum state space is spanned by linear combinations of arbitrary product of creation operators $a_T^\dagger$ and $a_L^\dagger$ multiplied onto $|0\>$.
The physical space is a part of this quantum space that fulfills eqs.(\ref{11}).
This space evidently has positive norm, which contrasts with the conventional quantization.

When we have interaction, the vacuum can be perturbatively obtained by \ben
	|\Omega \> &=& |0\> - \invHE V|0\>		\\
		&& + \invHE V \invHE V|0\> + \cdots.
\een
As was in section 1, we can confirm that $\phi^a|\omega\> = 0$ holds in the first order, where $\phi^a = \vec D \cdot\vec\pi^a  -\rho^a$.

Here in performing perturbation, $H_0$ is \[
	H_0 = {1\over 2} (\vec\pi^a)^2 + {1\over 2} (\vec\nabla \times\vec A)^2,
\]
and since $H = H_0 + V $, \be
	V = \int d^3x \left\{
		- {1\over 2} g f_{abc} (\vec\nabla \times\vec A^a) \cdot (\vec A^b\times\vec A^c) + {1\over 8} (g f_{abc} \vec A^b\times\vec A^c)^2
	\right\},
	\label{V3}
\ee
where $+A_0^a (\vec D\cdot\vec\pi^a - \rho^a)$ term in $V$ was ignored because this term does not have any effect as far as the constraints (\ref{11}) hold.
The constraint will further hold at any time because $H$ and the constraint commute owing to the consistency conditions.

Let us find the 1-particle state to use in the perturbative calculation in the next subsection, even though this state won't exist in a straightforward way thinking of confinement.
The perturbative 1-particle state is \be
	a_T^\dagger(\vec k)|0\> - \invHE V a_T^\dagger(\vec k)|0\> + \cdots,		\label{702}
\ee
where T in $ a_T $ stands for transverse polarization and its polarization vector will be designated by $\e_T$.
Its first-order part may be checked to fulfill the constraint.

\subsection{ Scattering }

Perturbative calculations of a scattering process will be discussed in this subsection.
Discussion of perturbation makes sense despite that perturbative states will not physically exist in a straightforward way because of confinement.
The first reason is that it is empirically known that gluon jet production at high energy agrees with QCD perturbation.
The second reason is that perturbative non-Abelian theory presented the problem of unitarity and lead to conventional formalism\cite{Fey63} of quantization of non-Abelian theory.  We must at least check that the unitarity problem is solved in our formalism.

Let us consider the process in that a lepton and an anti-lepton collide to produce two gauge bosons.
Each of the bosons will be experimentally observed as a jet.
The initial two particle state \[
	|\vec p_1, \vec p_2\>
\]
may be considered to be simply a product of 1-particle states since the infinitely dilute plane wave states have little interaction.
Here, we are going to calculate the lowest order contribution in $g$, which is the 2nd order, then we don't need to take the perturbative modification from the free 1-particle state $b_{p}^\dagger|0\>$.  Then we may calculate \ben
	\lefteqn{ e^{-iHT} b_{p_1}^\dagger b_{p_2}^\dagger|0\>}  			\\
	&=& \Big\{1+ \cdots + (-i)^2 \int_0^T dt_2 \int_0^{t_2} dt_1 e^{-iH(T-t_2)} V e^{-iH(t_2-t_1)} V e^{-iH t_1} \Big\} b_{p_1}^\dagger b_{p_2}^\dagger|0\>.
\een

The scattering amplitude may be obtained, in a manner similar to the QED scattering calculation, as \bea
	{\cal M'} &\equiv & \<0|a_{\sigma_1}(\vec k_1) a_{\sigma_2}(\vec k_2) e^{-iHT} b_{p_1}^\dagger d_{p_2}^\dagger |0\>		\nonumber\\
	&\equiv & (2\pi)^4 \delta^4(k_2 + k_1 - p_1- p_2) {1\over 4\sqrt{E_{p_1} E_{p_2} \omega_{k_1} \omega_{k_1}}}T_{ij}
\eea
by taking the leading contribution in the limit of the time interval $T$ getting large, where \bea
	T_{ij} &=& -i \bar v(\vec p_2) \Big\{ {\tau_b\over 2} \gamma_j {1\over  \p_1 - \k_1 - m }\gamma _i{\tau_a\over 2}
		 +\gamma _i {\tau_a\over 2} {1\over  \p_1 - \k_2 - m }{\tau_b\over 2} \gamma _j \Big\} u(\vec p_1)	\nonumber\\
		&& - f_{abc} V_{ij l} \Delta_{lk}(k_3)
				\bar v(\vec p_2) \gamma_k{\tau_c\over 2} u(\vec p_1).			\label{106}
\eea
Here $ k_{1\mu}+k_{2\mu}+k_{3\mu} = 0$ and \[
	V_{ijl}(k_1, k_2, k_3) = g_{ij}(k_1 - k_2)_l + g_{jl}(k_2 - k_3)_i + g_{li}(k_3 - k_1)_j.
\]
corresponds to eq.(\ref{V3}), which only have spatial components.

In this approximation, the L mode states won't appear in the final state.  L mode in this paper means the vibration mode (polarization) in which $ \vec A $ is proportional to $\vec k$.
The probability of appearance of one L-mode excitation is square of absolute value of the amplitude $ {\cal M'}$, and we calculate the following value to obtain it: \ben
	T_{ij} k_{2j} &=&  -i \omega\; \bar v(\vec p_2) \Big\{{\tau_b\over 2} \gamma_0{1\over  \p_1 - \k_1-m }\gamma_i {\tau_a\over 2}
		+\gamma_i {\tau_a\over 2}  {1\over  \p_1 - \k_2-m }{\tau_b\over 2} \gamma_0 \Big\} u(\vec p_1) 			\\
		&& - f_{abc} (k_{10}^2-k_{30}^2) \Delta_{ij}(k_3)
			\bar v(\vec p_2) \gamma_j {\tau_c\over 2} u(\vec p_1).
\een
In translating from eq.(\ref{106}) to this, we have used conservation of energy $ p_{10} - k_{10} = - p_{20} + k_{20},\; k_{20} = \omega$, and the zeroth component of the external momenta such as  $k_{10}, k_{20}$ is $\sqrt{\vec k^2+\omega^2} = \omega_k$ for transverse mode and $\omega$ for L mode.
We have also used that $ (\vec k^2 \delta_{ij} - k_i k_j ) \e_{j}(\vec k, \sigma) = (k_0^2 -\omega^2) \e_{i}(\vec k, \sigma),\; (\sigma = L, +, -) $ for external line, and $ (\vec k^2 \delta_{ij} - k_i k_j ) \Delta_{jk} = (k_0^2 -\omega^2) \Delta_{jk} $ holds for the photon propagator: \[
	\Delta_{ij}(k) = {i\over k^2 - \omega^2 + i\epsilon} \Big\{\delta_{ij} - {k_i k_j \over k_0^2 -\omega^2 +i \epsilon} \Big\}.
\]
Finally using $ k_{10}^2 - k_{30}^2 = \omega (k_{10} + k_{30})$, the amplitude becomes zero as \be
	\lim_{\omega\rightarrow 0} T_{ij} {\e_{j}(\vec k_2, L)  \over\sqrt{2\omega}} = 0.			\label{16}
\ee
Thus the probability of appearance of L mode state is zero.

On the other hand, if we don't take the approximation above, the L mode excitations actually appear and the electrostatic field of the final state particle consists of this excitation.
Here we consider the case when one of the two resultant gluons is L polarized.
By making similar discussion as in QED (subsection \ref{QED-L}), the final state $|\psi_{\rm f}\>$ with one L mode excitation is \bea
	|\psi_{\rm f}\> &=& \sum_{k_2} {g \over |\vec k_2|} J_0^a(-\vec k_2)  ({\vec \e(\vec k_2, L)}\cdot {\vec A^a(\vec k_2)}) |\psi_{\rm m} \>,			\\
%
%
	|\psi_{\rm m}\> &=& \sum_\sigma a_\sigma^\dagger(\vec p_1 + \vec p_2) |0\> \< 0|a_\sigma(\vec p_1 + \vec p_2) 
		\int^0_{-\infty} dy_0\; e^{+iHy_0} V e^{-iHy_0}\;
		b^\dagger_{p_1} d^\dagger_{p_2}|0\>,		\nonumber
\eea
where $J_\mu^a \equiv f_{abc} A^{\nu b} F_{\mu\nu}^c$. 
This final state $|\psi_{\rm f}\>$ agrees with the first order perturbative term in the 1-particle state (\ref{702}), i.e.\ this is a part of the 1-gluon state at the final time.
In this four-point case, this final state is that for the process $q +\bar q \rightarrow g$, which will only occur as an intermediate virtual state because of conservation of energy and momentum.  

\subsection{ Unitarity }

In our formalism, unitarity trivially holds for time evolution  $\exp(-iHt)$ because there is no negative norm states.
This contrasts with that it is a little bit complicated to show unitarity in currently prevailing BRST formalism.
Let us check it by an actual exemplar calculation.

The first non-trivial matrix element from the unitarity requirement is
\[
	2 {\rm Im} (-i) \<0|b_{p_1} b_{p_2} e^{-iHT} b_{p_1}^\dagger b_{p_2}^\dagger|0\>
	= \frac 1 2 \int {d^3k_1\over (2\pi)^3} \sum_{\sigma, \sigma'} \big|\<0|a_\sigma(\vec k_1) a_{\sigma'}(\vec k_2) e^{-iHT} b_{p_1}^\dagger d_{p_2}^\dagger |0\> \big|^2,
\]
and its lowest order is $g^4$ in the left-hand side and square of $g^2$ in the right-hand side.
This $g^4$ contribution in the left-hand side corresponds to 1-loop diagrams, and the right-hand side is 4-point scattering seen in the previous subsection.

Historically, it was considered that the states of gauge field theories consist of transversely polarized particles only, then the polarization sum in the right-hand side ran only over the transverse polarizations.  Then the right-hand side is \bea
	\frac 1 2 \int {d^3k_1\over (2\pi)^3} \sum_{{\rm color}; \sigma, \sigma' = \pm}
	T_{\mu\nu} \e^\mu(\vec k_1, \sigma) \e^\nu(\vec k_2, \sigma')
	T^*_{\mu'\nu'} \e^{*\mu'}(\vec k_1, \sigma) \e^{*\nu'}(\vec k_2, \sigma')
	{1\over 2\omega_{k_1}} {1\over 2\omega_{k_2}},			\nonumber\\
		\times (2\pi\delta(\omega_{k_1} + \omega_{k_2} - E))^2
	\label{old104}
\eea
where $T_{\mu\nu}$ is a expression obtained by replacing the 3-component indices $i, j$ with 4-component indices $\mu, \nu$ in eq.(\ref{106}), and $E = E_{p_1} + E_{p_2}$.
On the left-hand side, Feynman-gauge propagator was tried as the propagator of the virtual vector bosons for no reason since the correct way of quantization of non-Abelian gauge theory was not known at that time.
Therefore the left-hand side was considered to be \be
	{\rm Im} \; iT \int {d^4k_1\over (2\pi)^4} T_{ab\mu\nu} T^*_{ab\mu'\nu'} {g^{\mu\mu'} \over k_1^2} {g^{\nu\nu'} \over k_2^2}.
\ee
Performing $ k_{10}$ integration and taking imaginary part, we get \[
	\frac 1 2 T \int {d^3k_1\over (2\pi)^4} \sum_{{\rm color}}
	T_{\mu\nu} g^{\mu\mu'}
	T^*_{\mu'\nu'} g^{\nu\nu'}
	{1\over 2\omega_{k_1}} {1\over 2\omega_{k_2}}
	2\pi \delta(\omega_{k_1} + \omega_{k_2} - E) ,
\]
which does not agree with eq.(\ref{old104}).  This is the problem of unitarity, and lead to invention of the ghost particle by Feynman.

Contrarily, the polarization sum in eq.(\ref{old104}) runs over $\sigma, \sigma' = +, -, L$ in our quantization, and the left-hand side becomes \be
	{\rm Im} (-iT) \int {d^4k_1\over (2\pi)^4} T_{abij} T^*_{ab i'j'} \Delta^{i i'}(k_1) \Delta^{j j'}(k_2).
	\label{104}
\ee
Then, recalling that \[
	\Delta_{ij}(k) = \sum_{\sigma = \pm} {i\over k^2} \e_i(\vec k, \sigma) \e_j^*(\vec k, \sigma)
		+ {i\over k_0^2 - \omega^2} \e_i(\vec k, L) \e_j^*(\vec k, L),
\]
we integrate $ k_{10}$ and take imaginary part to get \bea
	\frac 1 2 T \int {d^3k_1\over (2\pi)^3} \sum_{{\rm color}; \sigma, \sigma' = \pm, L}
	T_{ij} \e_i(\vec k_1, \sigma) \e_j(\vec k_2, \sigma')
	T^*_{i'j'} \e^*_{i'}(\vec k_1, \sigma) \e^*_{j'}(\vec k_2, \sigma')			\nonumber\\
	\times {1\over 2\omega_{k_1,\sigma}} {1\over 2\omega_{k_2, \sigma'}} 2\pi \delta(\omega_{k_1,\sigma} + \omega_{k_2, \sigma'} - E),
\eea
where
$\omega_{k,\pm} = \sqrt{\vec k^2 + \omega^2}$,
$\omega_{k,L} = \omega$.
Now the right-hand side and the left-hand side agree taking $2\pi\delta(0) = \int_0^T dt = T$ into account. 

\subsection{ Loop calculation and Lorenz invariance }

Here we show an example of actual 1-loop calculations to exhibit easier way of  calculation and how we can make it Lorenz invariant form.
We take the approximation of taking leading contribution as $T$ goes large also in this subsection.

We first consider the amputated 1-loop 2-point graph of figure.1. \be
	\Gamma_2 \equiv
	{1\over 2} \int {d^4k_1\over (2\pi)^4} \; f_{acb} V_{\mu \rho \nu}(-k_1, k,-k_2) f_{bca} V_{\nu' \sigma \mu'}(k_2, -k, k_1)  \Delta^{\mu\mu'}(k_1) \Delta^{\nu\nu'}(k_2)			\label{10}
\ee
Here, the Roman indices $i, j,$ etc., which runs only spatial components, are replaced by Greek indices such as $\mu, \nu$ for the purpose of making covariance structure manifest.  Now $V$ is extended as \[
	V_{\mu \nu \lambda}(k_1, k_2, k_3) = g_{\mu\nu}(k_1 - k_2)_\lambda + g_{\nu \lambda}(k_2 - k_3)_\mu+ g_{\lambda\mu}(k_3 - k_1)_\nu.
\]
However, n-point Green functions won't be changed because the extended propagator will be defined as \[
	\Delta^{0 0}(k) = \Delta^{\mu 0}(k) = \Delta^{0 \mu}(k) = 0.
\]
Further, if any of the external indices $\rho, \sigma$ equal to zero, it may be ignored because the external indiceses of eq.(\ref{10}) is supposed be connected to $\Delta_{\rho\rho'}$.
Here each propagator $\Delta_{\mu\nu}(k)$ in the loop will be decomposed as
\bea
	-i k^2 \Delta_{\mu\nu}(k) = \delta_{ij} - k_i k_j/k_0^2
		= - g_{\mu\nu} + (k_\mu \eta_\nu+\eta_\mu k_\nu)/k_0 - k_\mu k_\nu/k_0^2,			\label{111}
\eea
where $\eta_\mu = (1,0,0,0)$.
Then we use \be
	V_{\mu \nu \lambda}(k_1, k_2, k_3) k_{1\mu} = (k_2^2 g_{\nu \lambda} - k_{2\nu}k_{2\lambda})-(k_3^2 g_{\nu \lambda}- k_{3\nu}k_{3\lambda}).			\label{wt1}
\ee
The $(k_3^2 g_{\nu \rho}- k_{3\nu} k_{3\rho})$ proportional term in this equation, which acts to the external line, will be discussed later and may be ignored in this paragraph.
We may ignore $V_{\mu \nu \rho}k_{1\mu}k_{2\nu} $ here since $(k_2^2 g_{\nu \rho}- k_{2\nu}k_{2\rho })$ times $k_{2\mu}$ is zero.
Then we have \ben
	\Gamma_2  &=&  \int {d^4k_1\over (2\pi)^4} V_{\mu\nu \rho} V_{\mu'\nu'\sigma} \Big\{ {1\over 2}g^{\mu\mu'} g^{\nu\nu'}
	 + (- k_{1\mu} \eta_{\mu'} - \eta_\mu k_{1\mu'})\frac 1{k_{10}} + k_{1\mu} k_{1\mu'} \frac 1{k_{10}^2} \\
	&& + (k_{1\mu} \eta_{\mu'}) (k_{2\nu'} \eta_\nu){1\over k_{10} k_{20}} \Big\} {1\over k_1^2 k_2^2},
\een
where we have abbreviated the momenta in $V$'s,
and used the symmetry between the two internal boson lines. Next using (\ref{wt1}), \ben
	\Gamma_2  &=&  \int {d^4k_1\over (2\pi)^4} \Big\{ {1\over 2}V_{\mu\nu \rho} V_{\mu'\nu'\sigma}  g^{\mu\mu'} 			\\
	&&
	- (k_2^2 g_{\rho\nu} - k_{2\rho}k_{2\nu}) {\eta_{\mu'} \over k_{10}}  V_{\mu'\nu'\sigma}
	- {\eta_\mu \over k_{10}} (k_2^2 g_{\sigma\nu} - k_{2\sigma}k_{2\nu'}) V_{\mu\nu \rho}				\\
	&& + (k_2^2 g_{\rho\nu} - k_{2\rho}k_{2\nu}) (k_2^2 g_{\sigma\nu} - k_{2\sigma}k_{2\nu'}) k_{2\mu} k_{2\mu'} \frac 1{k_{10}^2} 			\\
		&& + (k_{1\mu} \eta_{\mu'}) (k_2^2 g_{\rho\nu} - k_{2\rho}k_{2\nu})(k_{2\nu'} \eta_\nu)(k_1^2 g_{\sigma\nu} - k_{1\sigma}k_{1\nu'}){1\over k_{10} k_{20}}\Big\} {1\over k_1^2 k_2^2}.
\een
The terms containing $\eta_\rho$ for external indices $\rho, \sigma$ may be ignored, and here we may discard the terms proportional to $k_1^2$ or $k_2^2$ in the numerator since their results of integration are local terms, which can be removed by renormalization.
Then we have \be
	\Gamma_2  =  \int {d^4k_1\over (2\pi)^4} \Big\{ {1\over 2}V_{\mu\nu \rho} V_{\mu'\nu'\sigma} g^{\mu\mu'} g^{\nu\nu'}					 +  k_{1\rho} k_{2\sigma} \Big\} {i\over k_1^2} {i\over k_2^2}.			\label{607}
\ee
This expression is now Lorenz invariant and the effective action will be kept Lorenz invariant.
The second term agrees with the conventional Faddeev-Popov ghost contribution.  
We may perform renormalization as usual for such effective actions.
Wave function renormalization is \be
	Z_A = 1 + {g^2\over 16\pi^2} \times {10\over 3} N \log\Lambda.
\ee
Consequently, the 2-point function in the Feynman diagram will effectively behave \[
	{1\over k^2} \left(1 - {g^2\over 16\pi^2} \times {5\over 3}N \log {k^2 \over \mu^2} \right),
\]
where $\mu$ is the energy scale characteristic to non-Abelian gauge theory.

The terms that were ignored in the previous paragraph do not give any physical effect to n-point Green function.  They have the form of $(k^2 g_{\nu \rho}- k_\nu k_\rho)$ acting onto the external lines.
Let us consider the amplitude eq.(\ref{104}) in the previous subsection for example because cancellation is easier to see in actual physical process.
Here, $T_{ij}$ is also extended to have 4-component indices as $T_{\mu\nu}$.
The 2-point loop in the previous paragraph is included in this 1-loop amplitude(\ref{104}).
We have terms like $T_{\mu\nu} k_1^\mu$ if we use eq.(\ref{111}) to decompose $\Delta_{\mu\nu}(k)$, and it becomes
\bea
		T_{\mu\nu} k_1^\mu &=& -i \bar v(\vec p_2) \Big\{ {\tau_b\over 2} \gamma_\nu{1\over  \p_1- \k_1 - m } \k_1{\tau_a\over 2}
		 +\k_1 {\tau_a\over 2} {1\over  \p_1- \k_2-m } {\tau_b\over 2} \gamma_\nu \Big\} u(\vec p_1)	\nonumber\\
	&& + f_{abc}   (g_{\nu j} - \eta_\nu k_{3j}/k_{30}) \bar v(\vec p_2) \gamma_\lambda {\tau_c\over 2} u(\vec p_1)		\label{115} \\
	&& - f_{abc} (k_2^2 g_{\nu i} - k_{2\nu}k_{2i}) \Delta_{ij}(k_3)
		\bar v(\vec p_2) \gamma_j {\tau_c\over 2} u(\vec p_1).		\nonumber
\eea
The last term in the right-hand side corresponds to the contributions that was dealt with in the previous paragraph.
The first term cancels with the $ g_{\nu j}$ part in the second term using $ [\tau_a/2, \tau_b/2] = i f_{abc} \tau_c/2$.
The remaining part $k_{3\lambda} \bar v(\vec p_2) \gamma_\lambda(\tau_c/2) u(\vec p_1)$ is cancelled owing to the equations of motion for the fermions $(\p-m)u(\vec p) = 0$.
If this $k_\lambda$ proportional part acted on a external line,
it would become zero since it is orthogonal to $\e_\pm, \e_L$.
In summary, the terms in that $(k^2 g_{\nu \rho}- k_\nu k_\rho)$ acts onto a tree subdiagram 
will be cancelled by repeating the same procedure.

The results above are Lorenz invariant.
If we had quantized in a different Lorenz-boosted frame, we would simply have obtained the same formula but with boosted $ \eta = (1,0,0,0)$.
Physical results won't depend on the choice of the frame because the formula won't depend on $\eta$ as was seen in this subsection.


Let us next consider a 1-loop 3-point function shown in figure 2. \ben
	\int {d^4k_1\over (2\pi)^4} \; f_{dce} V_{\mu \gamma \nu'}(k_1, q_3, -k_2) f_{eaf}V_{\nu \alpha \lambda'}(k_2, q_1, -k_3)  f_{fbd} V_{\lambda \beta \mu'}(k_3, q_2, -k_1) 			\\
	\times\Delta^{\mu\mu'}(k_1) \Delta^{\nu\nu'}(k_2) \Delta^{\lambda\lambda'}(k_3)
\een
Using eq.(\ref{wt1}) similarly to the derivation of eq.(\ref{607}), ignoring the term with $k^2 g_{\mu\nu} - k_\mu k_\nu$ onto some external line, and considering about the cancellation with 1-loop diagram fig.(2b) with 4-point vertex, we obtain
\ben
	&&  - \int {d^4k_1\over (2\pi)^4} \; V_{\mu \alpha \nu'} V_{\nu \beta \lambda'} V_{\lambda \gamma \mu'} g^{\mu\mu'} g^{\nu\nu'} g^{\lambda\lambda'} {i\over k_1^2} {i\over k_2^2} {i\over k_3^2}			\\
	&& + \int {d^4k_1\over (2\pi)^4} \big\{ -  k_{1\gamma} k_{2\alpha} k_{3\beta}  -  k_{1\beta} k_{2\gamma} k_{3\alpha} \big\} {i\over k_1^2} {i\over k_2^2} {i\over k_3^2}.
\een
The first term agrees with the conventional result of vector field in the Feynman gauge, and the second term agrees with 1-loop contribution of the Faddeev Popov ghost.
The reason why these are the same may be understood by the discussion of gauge independence of FP determinant (subsection \ref{FPan}), but please remember that this is true under an approximation.		


\section{ Criticisms on conventional quantization methods }

In this section, we argue that all the conventional and common prescriptions of quantizing gauge field equally suffer from that the physical degrees of electrostatic field was eliminated.
This is because all the conventional prescriptions eliminated L mode, of which electrostatic field consists, as was seen in section 1 and 2.
We are going to explain what was wrong with each of conventional prescriptions in this section.

The widely used and acknowledged prescriptions are mainly:
\begin{itemize}
	\item gauge fixing by elimination (Coulomb gauge, axial gauge, etc.),
	\item Dirac's prescription,
	\item Gupta-Bleuler formalism,
	\item Feynman's argument (= belief of transversality) --- FP determinant --- BRST formalism,
	\item gauge fixing by adding constraints.
\end{itemize}
In this section, we will show that these are equivalent to each other except for that Gupta-Bleuler could not be used for non-Abelian theory.
It will be also shown that non-transverse polarization, i.e.\ L mode, is eliminated in all of them.
This section is for criticizing each of prescriptions concisely and we don't intend it as an introduction of the conventional methods.

\subsection{ Gauge fixing by Elimination }

The gauge fixing by elimination is an older prescription, in which gauge-fixing conditions is added in classical level to eliminate
impeding variables in moving to the canonical formalism.  Then we may move to canonical formalism and quantize.

In this prescription, we begin with the Lagrangian (e.g.\ eq.(\ref{LQED})) and variate $A_0$ to obtain one of the Euler-Lagrange equations $ \vec\nabla \cdot (-\dot{\vec A} - \vec\nabla A_0) = \rho$.
We may also derive this condition as an intermediate step eq.(\ref{12}) in moving to canonical formalism.

Then we add gauge-fixing conditions next.
For QED example, a popular choice is the Coulomb gauge condition $ \inp \nabla A = 0$.
Now we don't have the freedom of gauge transformation.
We now have a condition $ - \inp\nabla\nabla A_0 = \rho $ together with the other condition.
Only transverse vibration modes will be left if we eliminate $A_0$ from the Hamiltonian using this condition.  For the case of QED, the Hamiltonian is \[
	H_{0,{\rm trans}} + H_{\rm Coulomb},
\]
where $ H_{0,{\rm trans}}$ is a free Hamiltonian consisting of transverse modes only.  $ H_{\rm Coulomb}$ is the Coulomb energy shown in eq.(\ref{HCoulomb}), which is a bilocal function of the electron field  $\psi(x)$, i.e.\ electrostatic force is not the force from local field but now it is superluminal force with infinite reach.
It is possible to apply this prescription to non-Abelian theory in principle, but rarely applied because elimination using nonlinear relationships is cumbersome.
%

In this method above, elimination was carried out before completely moving to canonical formalism.
This may be said to be an elimination at Lagrangian stage, and two variables are eliminated in configuration space.
$A_0$ and $ \inp A \e_L =  A_L$ component, which were eliminated, won't become dynamical variables in the canonical formalism.
Elimination of two variables in the configuration space is elimination of 4 variables (freedoms) in the phase space.

Here, let us fabricate an intermediate method between ours and the above method to see what is happening in the above method of elimination.
We begin with eliminating $A_0$ from the Lagrangian using $A_0 = 0$, and we only impose $(\inp \nabla E - \rho)|*\> = 0$ onto every physical states instead of eliminating $A_L$ by $\inp \nabla E = \rho$.
Then we will have the same result as our theory.
That is, we obtain the same Hamiltonian as ours eq.(\ref{15}) except for $A_0(\dpi - \rho)$ term.
We remember that this term may be ignored due to the condition $(\inp \nabla E - \rho)|*\> = 0$.
Now we see that using $\inp \nabla E = \rho$ to eliminate is different from imposing $\inp \nabla E = \rho$ onto the physical states.
This difference can be understood as follows:
eliminating one variable before completely moving to Hamilton formalism, i.e.\ in configuration space, is eliminating two phase space variables or imposing two conditions onto the physical states;
however, imposing one condition, e.g.\ $\inp \nabla E |*\> = \rho$, onto the states after moving to canonical formalism is constraining one phase space variable.
Therefore, three degrees of freedom in the phase space are eliminated in this intermediate method.  This one degree of difference from the method of elimination corresponds to the degree of freedom for electrostatic field.
It is natural that $A_L$ component in configuration space corresponds to one degree of freedom in phase space because Euler-Lagrange equation for $A_L$, i.e.\ $ \inp \nabla E = \rho$, does not have 2nd order derivative in time.

On the other hand, only 2 degree of freedom in phase space was eliminated in our method in section 1.  The other freedom corresponds to the freedom of gauge transformation.
In other words, our method looks like $A_0 = 0$ gauge in the configuration space but different from it.
Because the states fulfill the condition $\pi_0 |*\> = \delta/\delta A_0 |*\> = 0$, we should rather think that all the possible configurations of $A_0$ are summed, i.e.\ all the possible gauges are summed.

\subsection{Dirac's theory}

The Dirac's canonical formalism of constrained systems
is classifying the constraints into ``the first class" and ``the second class", and defining ``Dirac brackets" instead of the Poisson brackets.
We will explain about it in this subsection.
How to do with the first class constraints will be commented later.
In section 1, we omitted explaining about the second class because only the first class constraints appear in the gauge theories.

Let us suppose that we have $N$ constraints $\{\phi_i \}$.
If a dynamical variable $R$ satisfies $ \{R, \phi_i\} \approx 0$ for all the $i$, this $R$ is defined to be ``the first class".
Unless, $R$ is ``the second class",
where $\approx$ is called ``weak equality" and
``$\approx$" means that ``equals when $\phi_i = 0\; (\forall i)$".
Each of constraints itself is also classified into the first or the second class.

Quantization was replacing Poisson brackets by commutators in non-constrained system, but it does not work for a constrained system with second class constraints.
Let us consider the simplest case of the 2nd class system $L =  x^2/2 $.
The primary constraint is $p = 0$ and the secondary constraint is $x = 0$ here.  If only one of them was required, there is no problem to realize it as well as eq.(\ref{5}) in section 1.  We can't, however, have $[x, p] = i\hbar$ when we have both.

Because of this problem, we can't replace Poisson brackets by commutators, and we have to define a new commutator that does not contradict the second class constraints.
The Dirac bracket was invented for this purpose, and replacing Dirac brackets with commutators is the Dirac's quantization for constrained system.
The Dirac bracket may be defined as \be
	\{F, G\}_D \equiv \{F, G\} - \{F,\phi_\alpha \} C^{-1}_{\alpha\beta}\{\phi_\beta, G\}
	\label{DirBr}
\ee
for the set of second class constraints\footnote{
When all the relevant constraints are second class, $ {\rm rank} C_{\alpha\beta} = N$ holds.
Unless we can make a linear combination of $\phi_\alpha$'s so that $C_{\alpha\beta} = 0$ for a column, and then that linear combination is a new first class constraint.
Here ${\rm rank}$ is defined on the subspace where all the constraints hold.
},
where $C_{\alpha\beta} \equiv \{ \phi_\alpha, \phi_\beta \} $.

Dirac himself proposed quantization by imposing \[
	\phi|{\rm phys}\> = 0
\]
for the first class constraints,
and using Dirac bracket for the second class\cite{Dir64}. 
Then it should have lead to the same argument as ours discussed in section 1 and 2.  The reason why he did not push this direction further is not known.
Please do not confuse the prescription discussed here with ``gauge fixing by adding constraints", which will be discussed later in this section, even though they look similar in that they both use Dirac brackets.

\subsection{ Gupta-Bleuler formalism }

Gupta-Bleuler\cite{GB50} formalism is a kind of method that uses subsidiary condition.
Subsidiary conditions are conditions that are imposed on the states to select physical states.
We need to explain about the difference between our method and this method to avoid confusion, because theirs is so much well known.

Gupta-Bleuler formalism is defined as imposing \[
	(\d A)_+|{\rm phys}\> = 0
\]
on the physical states, where $+$ means taking positive frequency part.
Further, they modify the Hamiltonian to be \[
	H = \sum_k E_{ k}( a_i^\dagger a_i- a_0^\dagger a_0)
\]
without good reason.
Because of this, the gauge degree of the freedom, which was originally arbitrary, and the electrostatic degree of freedom, which should not be vibrating, both vibrate with the frequency of $\sqrt{\vec k^2}$.

The reasons why Gupta-Bleuler (GB) formalism worked in QED are that the gauge degree of freedom did not make any effect because of Ward-Takahashi identity of electrons, and that L modes are suppressed by the subsidiary condition.
It is sometimes misunderstood that the subsidiary condition is for suppressing the gauge freedom; however, the gauge freedom is not suppressed since $\d_\mu \Delta A^\mu = k^2 \theta = 0$ owing to $k^2=0$ for gauge transformation $\Delta A_\mu= \d_\mu \theta$.
Instead, L mode is suppressed by this condition because $A_\mu \propto (0, \vec k)$.
The condition $k^2=0$ is because of the modification of the Hamiltonian.

For the reason above, L modes does not exist in GB formalism, there is no electrostatic freedom, and only transverse modes are left.
The theory only containing transverse modes does not make any problem as far as we are concerned with scattering in QED, and is equivalent to other theories in this section.

Note that GB formalism is different from other conventional methods in this section in that it does not eliminate gauge freedom, and just relies on Ward-Takahashi identity $\d_\mu j^\mu = 0$ on the electron side.
This is the reason why Gupta-Bleuler formalism did not work for non-Abelian gauge theory unlike the other methods.

\subsection{Feynman's ghost and Faddeev-Popov ansatz}

Feynman started with the idea that only transverse polarization is possible for physical states in non-Abelian gauge theory\cite{Fey63}.
Then unitarity was violated in 1-loop order, and he invented a fictitious particle, which is called ``ghost",  to make the theory consistent.
This theory also is equivalent to other already presented methods, which have only transverse modes.

This Feynman's theory was generalized to higher-loop order by DeWitt\cite{DeWitt}, and Faddeev and Popov\cite{FP} had shown that a simple rule reproduces that result.

The path-integral formula for the gauge field is \[
	\int {\cal D}A\; O_1 O_2 \cdots \exp i\int d^4x\; {\cal L}(x) ,
\]
where $O_1 O_2 \cdots$ is some combination of fields, and ${\cal L}(x)$ is the Lagrangian density.
However, this expression cannot be calculated, because it has infinite volume where ${\cal L}(x)$  has the same value because of gauge independence.
To solve this problem, Faddeev and Popov proposed replacing that formula by \bea
	I = \int {\cal D}A\; O_1 O_2 \cdots   \Delta[A,f] \prod_{x, a} \delta (f[A^a(x)])
		\times \exp i\int d^4x\; {\cal L}(x),				\label{21}
\eea
where $f$ is a functional of $A^a(x)$.
This expression does not depend on the condition $f[A(x)] = 0$, which pins down the gauge arbitrariness.
Here, \bea
	\Delta[A, f] &\equiv& \left( \int {\cal D}\theta \prod_{x,a} \delta(f[A^{\theta a}(x)]) \right)^{-1},			\label{24} \\
	A^{\theta a}(x) &=& \d_\mu \theta^a(x) + i g f_{abc} A_{\mu}^b(x) \theta^c(x),		\nonumber
\eea
where ${\cal D}\theta$ means gauge independent functional integral on 4-dimensional space-time.

Next, let us prove that eq.(\ref{21}) is independent of choice of $f$.  At first, $\Delta[A, f]$ is independent of gauge transformation \[
	\Delta[A^\theta, f] = \Delta[A, f]
\]
since ${\cal D}\theta$ is invariant measure against gauge transformation.  Bringing a different $ f' $ into eq.(\ref{21}), we may rewrite it as \bea
	I &=& \int {\cal D}A\; O_1 O_2 \cdots \Delta[A,f] \left[ \prod_{x,a} \delta (f[A^a(x)]) \right] 			\nonumber\\
		&&\times \Delta[A, f'] \int {\cal D}\theta \prod_{x,a} \delta(f'[A^{\theta a}(x)])
	\exp i\int d^4x\; {\cal L}(x).	\label{22}
\eea
By letting the ${\cal D}\theta$ integration in eq.(\ref{22}) outermost and using gauge invariance of ${\cal L}$ and $\Delta[A, f]$, \ben
	I &=&\int {\cal D}\theta\int {\cal D}A\; O_1 O_2 \cdots \Delta[A^\theta,f] \left[ \prod_{x, a} \delta (f[A^{\theta a}(x)]) \right] 			\\
		&&\times \Delta[A, f'] \prod_{x,a} \delta(f'[A^a(x)])
	\exp i\int d^4x\; {\cal L}(x).	
\een
Therefore, we could show \[
	I = \int {\cal D}\theta \int{\cal D}A\; O_1 O_2 \cdots  \Delta[A, f'] \prod_{x,a} \delta(f'[A^a(x)])
		\times \exp i\int d^4x\; {\cal L}(x).
\]
In other words, this formula does not depend on $\delta(f[A])$ factor, which pins down the gauge arbitrariness, and this is the excuse to introduce eq.(\ref{21}).

Let us now evaluate $\Delta[A,f] $.
$\Delta[A, f] $ may be evaluated assuming $f[A](x) = 0$ because we have $\delta(f[A])$ in eq.(\ref{21}).
Then eq.(\ref{24}) may be evaluated around $\theta(x) = 0$,
and the gauge independent integration $\int {\cal D}\theta$ becomes simply $\int \prod_a d\theta ^a$, and finally eq.(\ref{24}) is equal to \ben
	\int \prod_a d\theta^a \prod_{x,a} \delta\left( \left. {\delta f[A^{\theta a}] \over \delta \theta^b(x)} \right|_{\theta=0} \!\cdot\theta^b(x) \right)
		= \det \left.{\delta f[A^{\theta a}] \over \delta \theta^b(x)} \right|_{\theta=0}.
\een
Especially, when we take Landau gauge $ f[A] = \d A = 0$,
\[
	\Delta = \det \d_\mu (\d^\mu \delta _{ab}+f_{abc} A^{\mu c}).
\]
This is called Faddeev-Popov determinant, and its contribution to $I$ coincides with that of Feynman's ghost particle at 1-loop.

\label{FPan}

On the other hand, if we take gauge fixing by $A_0 = 0$ in the configuration space instead of the Landau gauge, the determinant to appear is \[
	\Delta^{-1} = \int {\cal D}\theta \prod_{x,a} \delta(A_0^a(x)) = \det 1.
\]
Then \[
	I = \int {\cal D}\theta\int {\cal D}A\; O_1 O_2 \cdots \prod_{x,a} \delta(A_0^a(x))
	\times \exp i\int d^4x\; {\cal L}(x)
\]
This will give the same value as ours as already mentioned.
It has been considered that changing gauge-fixing condition does not change physical results by the ``proof" already presented; however,
our theory is different in that it has state that corresponds to electrostatic field.
\delete{
the fact that the three-point function in section 2 is different from the conventional one is a counterexample for that.
It is probable that this failure of independence from gauge-fixing condition is because of the problem that was pointed by Gribov\cite{Grib}.
The number of the solutions of gauge fixation condition $f[A^\theta](x) = 0$ can depend on the configuration $A(x)$ and the gauge condition $ f $.
The aforementioned ``proof" about independence did not take this copy number into account.
}

\subsection{ BRST formalism }

The FP determinant in covariant gauge $\det \d_\mu D^\mu$ may be written as an path-integral of Grassmann fields as \ben
	&& \det\d^\mu D_\mu = \int Dc D\bar c\; \exp i\int d^4x\; [i \bar c^a \d^\mu D_\mu c^a],			\\
	&& D_\mu c^a = \d_\mu c^a + g f_{abc} A_\mu^b c^c.
\een
This fields $c, \bar c$ are called ``Faddeev-Popov (FP) ghost" regarding it as an particle.
We rewrite the factor that was fixing gauge in eq.(\ref{21}) as \[
	\prod_{x, a} \delta(\d_\mu A^{a \mu})
	 = \int {\cal D}B\; \exp i\int d^4x \{ B^a (\d_\mu A^{a\mu}) + \alpha B^2/2\},
\]
where we have taken the Landau gauge for brevity, which requires $\alpha=0$.
Including these FP ghost and gauge-fixing terms, the total Lagrangian density becomes \[
	\tilde{\cal L}(x) \equiv {\cal L}(x) + B^a (\d_\mu A^{a\mu}) + \alpha B^2/2 + [i \bar c\; \d^\mu D_\mu c].
\]

This $\tilde{\cal L}$ with FP ghost is invariant under a peculiar transformation, which is called ``BRST transformation":
\bea
	\delta A_\mu^a &=& \d_\mu c^a + g f_{abc} A^b_\mu c^c,		\nonumber\\
	\delta c^a &=& - g f_{abc} c^b c^c /2,			\label{BRST}	\\
	\delta\bar c^a &=& i B^a,			\nonumber\\
	\delta B^a &=& 0.			\nonumber
\eea
Operating this transformation twice will always give zero.
The charge for this transformation is called ``BRST charge" $Q_B$, and it have properties of $ Q_B^2 = 0$ and $[Q_B, H] = 0$.

It is shown by Kugo and Ojima\cite{KugOji} that $Q_B |\psi\> = 0$ if a state $|\psi\>$ consists only of transverse gauge bosons, and conversely, $|\psi\>$ consists only of transverse states and zero probability states if $Q_B |\psi\> = 0$.

If the initial state consists of transverse polarization only, it is true all the time, which can be shown from this theorem.
That is because $Q_B |*\> = 0$ at the initial time, and then $Q_B |*\> = 0$ all the time since $ H $ and $ Q_B $ commute.

\delete{
Though it is hard to tell if FP formalism contains only transverse mode or not, but now we see that its rewritten form, i.e.\ BRST formalism, contains transverse modes only.
}
In other words, this formalism also wipes out L modes and has no electrostatic field.

\subsection{ Gauge fixing by Adding Constraints }
\label{addingconstr}

The Dirac bracket can't be defined if we include any first class constraint $\phi_a$ because $C_{ab} \approx 0$, and ${\rm rank} C \neq N$.
Then all the constraint will be rendered second class by adding $N - {\rm rank} C$ constraints ${\xi_b}$ by hand\footnote{It is supposed that the equations of motion from the Hamiltonian have $N - {\rm rank} C$ degrees of arbitrariness.}.
This is the prescription of gauge fixing by adding constraints.
We choose ${\xi_b}$ so that the determinant of \be
	\left( \begin{array}{cc}
	\{\phi_\alpha, \phi_\beta \} & \{\phi_\alpha,\xi_\beta\}			\\
	\{\xi_\alpha, \phi_\beta \} & \{\xi_\alpha, \xi_\beta \}
	\end{array}\right)			\label{17}
\ee
is not $\approx 0$.
Then the problem is reduced to that of a system with the second class constraints only, which may be quantized using Dirac brackets.
Different gauge fixating conditions all give the same result because it will be shown to be equivalent to Faddeev-Popov prescription, which is said to be gauge invariant.

Let us apply this prescription to the actual case of non-Abelian gauge field.  For example, let's take a common choice of the gauge \bea
	\xi_1^a &=& A_0^a \approx  0			\nonumber\\
	\xi_2^a &=& \inp \nabla  A^a \approx  0.			\label{18}
\eea
This choice gives the following Poisson brackets: \be
	\{\xi_i^a, \phi_j^b \}
	= \left(\begin{array}{cc}
		\delta^{ab}\delta^3(\vec x - \vec y)	&	0 			\\
		0	&	\vec\nabla _x \cdot (\delta^{ab} \vec\nabla _x - g f_{abc} A^c(x))\delta^3(\vec x - \vec y)
	\end{array}\right)
	\equiv M(t)
	\label{Mt}
\ee
Then the determinant of matrix (\ref{17}) is not zero.
Owing to this, we can define Dirac brackets and can quantize by replacing them with commutators.

The above result is different from ours.
Especially, the condition $\vec\nabla \cdot\vec A^a \approx  0$ means that  $ A_\mu \propto (0, \vec k)$ part, i.e.\ L mode, is constrained not to appear.
Then electrostatic field, which depends on L mode, won't exist.
This prescription is same as the Coulomb-gauge elimination in that only transverse modes are left.

To make this addition of constraints valid, the canonical variables should have as many freedom as the number of constraints to be added.
In the current example of gauge theory, we need two degrees of freedom for each space-time point.
Though, we only have one degree of freedom of $\lambda$ for each space-time point as far as we see equations of motion from Hamiltonian (\ref{HnA}).
Further, we have another question.  In Coulomb gauge in the method of elimination, the electrostatic field was given by $-{\vec\nabla}^2 A_0 = \rho$.  How can we have electrostatic field when $A_0 = 0$ and $A_L = 0 $?

The answer is that it was considered possible to add two conditions without changing dynamics because of Dirac's extended Hamiltonian $H_E$.
According to Dirac, one should generalize Hamiltonian by adding not only the primary constraint $\pi_0$ but also the first-class secondary constraint $\dpi -\rho$ as \be
	H_E = H + \lambda_1(\vec x,t)\pi_0 + \lambda_2(\vec x,t) (\inp \nabla \pi-\rho ),			\label{19}
\ee
which add two arbitrariness for each space-time point.
Dirac justified adding it because it won't change any state in the classical sense\cite{Dir64}. 
However, the constraints and equations of motion that is derived from $H_E$ are
\bea
	\pi_0 &=& 0,		\nonumber\\
	\dpi  &=& \rho,			\nonumber\\
	\dot{\vec\pi} &=& \vec\nabla \times\vec\nabla \times\vec A,			\\
	\dot{\vec A} &=& - \vec\pi - \vec\nabla A_0 - \vec\nabla \lambda_2,			\nonumber\\
	\dot A_0 &=& \lambda_1.			\nonumber
\eea
Here $\lambda_1$ is arbitrary and gives correct gauge transformation also to $\vec A$ through $A_0$.
On the other hand, the change caused by $\lambda_2$ modifies Euler-Lagrange equation in the fourth equation, and this change is not a gauge transformation, which is hard to justify.
Since the $\vec k$ proportional part $\pi_L$ is determined by the second equation, the $\lambda_2$ changes $A_L$.
Especially under $A_0 = 0, A_L = 0$ condition, electrostatic field is supported by $\lambda_2$ as $\vec E = - \vec\nabla \lambda_2$. 
Here we again don't have any dynamical variable that corresponds to electrostatic field since $\lambda_2$ is not a quantum variable.

\subsection{ Equivalence of quantized second class constrained system and Faddeev-Popov ansatz }

There is a theorem that the Dirac bracket method for a system with only second class constraints is equivalent to solving constraints, and quantization after this elimination gives a simple path-integral formula.
Using these theorems, it will be shown that FP ansatz is equivalent to quantization with added constraints (subsection \ref{addingconstr}).

Let us call the phase space spanned by the $2N$ variables $p_i, q_i$ by $\Gamma$.
We suppose that we have constraints $\phi_j = 0 \; (j=1, \cdots, 2m)$ that specify a subspace $\Gamma^*$ in $\Gamma$.
Then we can find a set of canonical variables $p^*_i, q^*_i (i=1,\cdots, N-m)$ that can be coordinate variables for $\Gamma^*$.
In other words, we can solve $\phi_i$ to have $2N-2m$ dimensional canonical variables.
Moreover, the Dirac brackets are, in fact, the Poisson bracket in $p^*_i, q^*_i$, i.e.\ \be
	\{ A, B \}_D = \sum_{i=1}^{N-m}
		\left\{ \frac{\d A}{\d q^*_i} \frac{\d B}{\d p^*_i}
		- \frac{\d A}{\d p^*_i} \frac{\d B}{\d q^*_i} \right\}.
\ee
These theorems were proven by Maskawa and Nakajima\cite{MasNak76}.

If we quantize solved variables $p_i^*, q_i^*$ in $\Gamma^*$, the path-integral formula is \[
	\int d^{2N-2m}p^*\; d^{2N-2m}q^* \; O_1 O_2 \cdots \exp i\int dt [p^*\dot q^* - H(p^*, q^*)].
\]
This formula equals to \bea
	\int d^{2N}p\; d^{2N}q \; O_1 O_2 \cdots \Big[ \prod_t \prod_{\alpha = 1}^{2m} \delta(\phi_\alpha) \Big] (\det\{\phi_\alpha, \phi_\beta \})^{1/2}			\nonumber\\
		\times \exp i\int dt [p \dot q - H(p, q)],			\label{20}
\eea
which was also proven in the paper \cite{MasNak76}.

We can show that Dirac bracket method for second class constraints is equivalent to Faddeev-Popov method using these theorems.
Here we use gauge-fixing condition (\ref{18}) for example, and recall the matrix of Poisson bracket (\ref{Mt}) among the constraints.
Substituting $\det M$ of this matrix into (\ref{20}), we have
\ben
	\int {\cal D}\pi {\cal D}A\; O_1 O_2 \cdots \left[ \prod_{x,a} \delta(A_0^a(x)) \delta(\inp \nabla A^a(x))				\delta(\pi_0^a(x))\delta(\Dpi^a(x)-\rho^a(x)) \right] \\
	\times \prod_t \det M(t)\; \exp i\int d^4x [\pi^\mu\dot A_\mu - H(p, q)].
\een
Then we integrate $A_0$ and $\pi_0$ to get \ben
	\int {\cal D}\pi {\cal D}A\; O_1 O_2 \cdots \left[ \prod_{x,a} \delta(\inp\nabla A^a(x)) \delta(\Dpi^a(x)-\rho^a(x)) \right] \prod_t \det M(t)			\\
	\times \exp i\int d^4x [\pi^\mu\dot A_\mu - H(p, q)],
\een
and introducing a different $A_0$ and using \ben
	 \prod_a \delta(\Dpi^a(x)-\rho^a(x))
	 = \int {\cal D}A_0 \exp i\int d^4x [A_0^a (\Dpi^a - \rho^a)],
\een
and integrating $\vec\pi$, eq.(\ref{20}) finally becomes
\bea
	\int {\cal D}A\; O_1 O_2 \cdots \left[ \prod_{x, a} \delta(\inp\nabla A^a(x)) \right] \prod_t \det M(t)
		\times \exp i\int d^4x \; {\cal L}(x).			\label{21b}
\eea
Here we could show that Faddeev-Popov determinant was reproduced because this equation (\ref{21b}) is eq.(\ref{21}) substituted by $ f = \inp\nabla A$.

\subsection{ Summary }


In this section, we have shown that all the widely-used formalisms had eliminated electrostatic field.
In these formalisms, not only the freedom of gauge transformation but also the freedom of electrostatic field was eliminated because of gauge fixing, contrary to ours.

The conventional methods each had different reason for justification.
Gauge fixing by elimination was justified by thinking that variables may be eliminated before quantization.  They thought that two variable may be solved away even though the gauge transformation has only one degree of freedom.  It was possible because one of the conditions is the Euler-Lagrange equation related with electrostatic field.  Then the electrostatic field was solved away classically.
Feynman's ghost justified itself by making the idea self-consistent that only transverse polarization is physical.
FP ansatz was justified by that path-integral must be gauge invariant, and it was the only known way.
Gauge fixing by adding constraints was justified by thinking that 4 variables may be eliminated before quantization.
They think that 4 variable may be solved away even though the gauge freedom is only one degree.  The elimination was possible because Dirac's extended Hamiltonian increased the freedoms and a new undetermined variable took over the electrostatic field.

Why all the conventional methods consistently lead to the same result, even though they each had different reason for justification?
The suspected reason is that they each constructed these theories by giving glance on the quantum theory of radiation.
The belief of transversality prevailed in the quantum theory of radiation from its beginning, i.e.\ in 1930's.

These conventional theories are not wrong by itself, because they are self-consistent.
The question is which of ours or theirs does describe the physics more appropriately.

\section{ Discussion and Conclusion }

In this paper, the belief of transversality was discredited.
The belief of transversality is that only transverse polarizations are physical degrees of freedom and need to be quantized.
If one believe that only transverse modes exist, it well lead to non-existence of electrostatic field as quantum object.
Moreover, our formalism has shown it possible to describe a configuration of electrostatic field as a quantum state.
The observable of electrostatic field obtained from our theory took continuous value.

In our intuitive understanding, electrostatic field is a measurable quantity that exists at every space point, and every measurable quantity should have a quantum counterpart as an observable.
In the conventional theory, in contrast, the electrostatic field is not a physical observable on space but a superluminal force between electrons, which is easily seen in the Coulomb gauge.
This is counterintuitive but thought to be inevitable.  Our formalism allows intuitive and natural understanding on the other hand.

The reason why this problem of missing electrostatic field was not recognized is that field theories are used mainly in scattering calculation and then electrostatic field did not matter.

Many people believe that reduction commutes with quantization.  That is, we will get the same theory if we first eliminate classical gauge freedoms and then quantize, or if we quantize and then take gauge-invariant subspace.  A mathematical counterpart of this question is called Guillemin-Sternberg conjuecture\cite{GuilSter}, and proven in a limited setup of finite dimension.
The method of elimination and the method of adding constraints are examples of this classical reduction, and they are different from ours.  So, this conjecture should not hold in gauge field theories.
The reader may also have the same idea as the author that Dirac has proposed this equivalence under the commutation.  Actually, Dirac did for the second class constrains, but not for the first class constraint. 

It seems that our formalism have a problem of less looking covariant in spite of its advantages.
Because we have the condition $\pi_0 |*\> = \delta/\delta A_0 |*\> = 0$ on the states, we should rather understand that all the possible configurations of $A_0$ are summed, i.e.\ all the possible gauges are summed in our formalism.
Since it is the sum of all the possible configurations, we may think it is totally covariant overall.
The reason why it looks non-covariant may be that the method for solving the equations of motion is dependent on the time slice, e.g.\ the mode expansion formula (\ref{A}) is dependent.
Further, we can ensure existence of solution and don't have multiple solutions for the equation $A_0(x) = U^\dagger(x) \d_0 U(x)$, which is necessary for gauge transforming from arbitrary $A_\mu(x)$ to $A_0 = 0$.
This is better than the other gauges, e.g.\ $\d_\mu A^\mu = f$, for which neither existence nor multiplicity of solution is known well.


Next, let us comment about implications of our theory.

The dynamics of color-electric field line should be the cause of confinement of quarks in QCD, therefore the conventional quantization methods, which has only with transverse mode, won't be able to show confinement.  
In contrast to the conventional theory, our theory may predict such thing as the separation dependence of color electric force between two stationary quarks. %
Showing the mass gap and confinement in the non-Abelian gauge theory is a major problem in physics and mathematics.
The ground, or the axiom, on which the theorem of confinement and mass gap should be proven, is now changed.

In the attempts at quantizing gravity, we frequently come across upon the problem of whether we could deduce ordinary continuous space-time or not.
Since the parallel between electrostatic fields and the continuous static space-time is apparent, we may expect deducing continuous space-time likewise.
It may be reasonably suspected that the belief of transversality is the cause of the difficulty of the current attempts at quantum gravity too.


The conventional formalisms, especially BRST formalism, are eliminating not only the freedom of gauge transformation but also the freedom of electrostatic field.
In other words, the freedoms of FP ghost is equal to the freedom of gauge plus that of electrostatic field, i.e.\ L modes.
We may also say that the BRST symmetry is a two-fold extended symmetry that consists of not only gauge symmetry but also forcefully added freedom of electrostatic field.

The theorems that were proven using the BRST symmetry will need rework.
If a theorem is not wrong physically or mathematically, it will be possible to reinterpret the theorem in physically meaningful way.
It is likely that the theorems will become the ones about electrostatic field because BRST symmetry minus the trivial gauge symmetry is electrostatic part.


Let us next comment about implication on the Higgs mechanism.
Because there is no helicity zero degree of freedom in the conventional formalisms, we must supply degrees of freedom from elsewhere to make a massive vector boson possible.  For this reason, the Higgs particles were required.
But in our case, we have helicity zero states, that is the L modes, and then it may be possible to have massive vector boson without Higgs particle.  This possibility was already discussed in \cite{MG} by the author. 


The lattice gauge theory should be equivalent to our theory because, in dealing with gauge symmetry, they are the same in that they both don't do forced gauge-fixing, and that no kinetic term exists for L mode, i.e.\ zero energy excitement.
It will soon be shown in the numerical experiments that the lattice gauge theory agrees to ours.


In conclusion, our formalism suggests another consistent way of quantizing gauge theory.
The BRST formalism and the belief of transversality should need careful revision, where they are closely related to each other as explained in section 3.
So, all the works that rely on the belief of transversality will need rethink.

\end{document}